\documentclass[12pt,preprint]{aastex}
\usepackage[dvips]{color}

\slugcomment{Fermilab-PUB 02-036/A; CERN-TH/2002-032}

\shorttitle{Stellar Mass Black Holes}
\shortauthors{Chisholm, Dodelson, Kolb}

\begin{document}

\title{STELLAR-MASS BLACK HOLES\\ IN THE SOLAR NEIGHBORHOOD}

\author{\sc James R. Chisholm}
\affil{Theoretical Astrophysics,
Fermi National Accelerator Laboratory, Batavia, Illinois 60510}
\affil{Department of Physics, The University of Chicago,
Chicago, IL 60637}
\email{chisholm@oddjob.uchicago.edu}

\author{\sc Scott Dodelson}
\affil{Theoretical Astrophysics,
Fermi National Accelerator Laboratory, Batavia, Illinois 60510}
\affil{Department of Astronomy and Astrophysics, The University of Chicago,
Chicago, IL 60637}
\email{dodelson@fnal.gov}

\and

\author{\sc Edward W. Kolb}
\affil{Theory Division, CERN, CH-1211 Geneva 23, Switzerland}
\affil{Theoretical Astrophysics,
Fermi National Accelerator Laboratory, Batavia, Illinois 60510}
\affil{Department of Astronomy and Astrophysics, The University of Chicago,
Chicago, IL 60637}
\email{rocky@fnal.gov}

\begin{abstract}

We search for nearby, isolated, accreting, ``stellar-mass'' ($3$ to
$100M_\odot$) black holes.  Models suggest a synchrotron spectrum in visible
wavelengths and some emission in X-ray wavelengths.  Of 3.7 million objects in
the Sloan Digital Sky Survey Early Data Release, about 150,000 objects have
colors and properties consistent with such a spectrum, and 87 of these objects
are X-ray sources from the ROSAT All Sky Survey.   
Thirty-two of these have been 
confirmed not to be black-holes using optical spectra.  We give the positions 
and colors of these 55 black-hole 
candidates, and quantitatively rank them on their likelihood to be black holes.  
We discuss
uncertainties the expected number of sources, and the contribution of black
holes to local dark matter.
\end{abstract}


\section{INTRODUCTION}

Since they were first postulated, black holes have captured the imagination
of scientists, as well as the general public. In addition to their intrinsic 
appeal, black holes potentially
impact on a number of fundamental problems in physics and astronomy.  They are
a possible endpoint of stellar evolution.  They provide a unique laboratory in
which to study strong gravity. Knowledge of the mass and spatial distributions
of black holes could also provide information about stellar evolution, galaxy
formation, and dark matter.

While they are among the most interesting astrophysical objects, black holes,
by their very nature (``black''), are difficult to isolate and study.  Since
the intrinsic Hawking radiation from black holes of the mass we study (greater
than about a solar mass) is quite feeble, the search for black holes must
concentrate on the interaction of the black hole with the surrounding medium.
Thus, the search strategy for black holes depends upon the hole mass.

We can organize black holes in groups according to their mass.  Super-massive
black holes with mass of order $10^8\,M_\odot$, are thought to reside in the
nuclei of galaxies [for a review, see \citet{Kormendy}].  In addition to their
role in the dynamics of galaxies and galaxy formation, they are believed to be
the central engines of energetic phenomenon associated with active galactic
nuclei. 

Evidence for intermediate-mass black holes of around $10^2$ to $10^5\,M_\odot$
has recently been found \citep{Matsumoto99, Ptak, Matsumoto01,
Kaaret}. Intermediate-mass black holes might be precursors to super-massive
black holes \citep{Ebisuzaki}.

Lower mass black holes formed at the endpoint of stellar evolution are known as
remnant black holes (RBH).  They are expected to have masses from about $3$ to
$100M_\odot$, and are the focus of this paper.  These remnant black holes are
sometimes referred to as ``stellar-mass'' black holes.  The lower mass limit
derives from the upper mass limit of neutron stars, the other possible outcome
for the core of a supernova.  The upper mass is limited by the mass of the
progenitor star and possible subsequent accretion onto the RBH.  Remnant black
holes in binary systems were first discovered through X-ray emission [see,
e.g., \citet{Tanaka}].  The first observational hint for the existence of an
{\it isolated} RBH comes from gravitational microlensing \citep{Bennett, Mao}.

A last group of black holes, which would have masses less than $3M_\odot$,
could not have been formed through stellar evolution, but might have formed at
an earlier time in the universe, presumably as primordial black holes.  

Of the four groups, there is good evidence for the existence of the first three
types of black holes.  This paper discusses a search for candidates for {\it
isolated} remnant black holes.

As mentioned, since black holes will not themselves be luminous, the key to
detecting them is to observe their effect on their surroundings.  Black holes
will accrete and radiate some fraction of the accreting mass into energy.
[For a review of black-hole accretion, see \citet{Chakrabarti}.]

For those RBHs that are part of a binary system, accretion is typically from
the companion star through Roche lobe overflow onto an accretion disk.  These
objects were first discovered through their X-ray emission \citep{Tanaka},
though detections have been made from radio to $\gamma$-ray frequencies.  The
accretion emission in the optical and near-infrared is dominated by that of the
companion star.

For isolated RBHs,\footnote{Henceforth, unless specifically indicated, when we
refer to a remnant black hole, it should be understood that we mean an {\it
isolated} remnant black hole.} the subject of this study, the accretion is from
the interstellar medium. The accretion rate from the ISM is typically much less
than that from a companion star, so the corresponding luminosity of an isolated
RBH is so small that they have yet to be detected.  Since nearly half of all
stars are in binary systems, we might expect the RBHs formed to have this same
ratio.  Thus, isolated RBHs should outnumber binary RBHs by 2:1.

Accretion emission is not the only proposed method for detecting RBHs.  There
is recent evidence \citep{Bennett,Mao} for the detection of black holes using
gravitational microlensing.  Another method is looking for the effect of black
hole creation in supernova light curves \citep{Balberg}. Finally, another
method for finding a RBH is to look for the cutoff of neutrino emission due to
black hole formation during a supernova \citep{Beacom}.

The purpose of this paper is to identify a small number of isolated, nearby,
RBH candidates for follow-up optical and X-ray studies.  Our approach,
following the suggestion of \citet{Heckler}, is to search for objects in the
Early Data Release \citep{Stoughton} of the Sloan Digital Sky Survey (SDSS)
\citep{York} that are consistent with 
the type of power-law spectra expected
from isolated RBHs\footnote{Our starting point is the SDSS optical survey.
Recently \citet{Agol} have discussed a X-ray based search strategy.}. We also
require that the source appear as an X-ray source in the ROSAT all-sky survey
\citep{Voges}.  

\citet{Heckler} suggested using the SDSS spectroscopic survey to detect RBHs. 
The photometric survey however is four to five magnitudes fainter so is
sensitive to objects ten times further away, and will have a thousand times
more objects than the spectroscopic survey.  A possible disadvantage of the
SDSS photometric survey is that it lacks detailed spectral information, making
it more difficult to identify positively RBH candidates while simultaneously
rejecting spurious stars. However, we find that the five color bands of the
SDSS photometric survey are quite adequate for synthesizing the optical
spectrum, and the cross-correlation with ROSAT eliminates the vast majority of
stars.

In the following section we discuss theoretical issues and uncertainties
involving the expected spectrum, the expected RBH mass and velocity
distributions, and estimates of the number densities of RBHs.

In \S 3 we describe the search strategy, and \S 4 contains our results. We
conclude in \S 5.  The major result are tables of positions and color
magnitudes of the RBH candidates.

\section{THE SPECTRAL ENERGY DISTRIBUTION FOR ISOLATED REMNANT BLACK HOLES}

In this section we discuss possible spectral energy distributions for isolated
remnant black holes.  The spectrum depends upon the mass of the hole and the
local environment of the RBH (the density of the interstellar medium (ISM), and
the velocity of the hole with respect to the ISM).  Even with knowledge of the
local conditions surrounding the RBH, there is still a great deal of
uncertainty in the spectral energy distribution.

We can roughly classify RBH accretion models by geometry (spherical or
disk-like) and type (thermal, advection-dominated, convection-dominated,
etc). Most, if not all, accretion models utilize magnetic fields in one way or
another.  In all models, it is the resulting electron synchrotron radiation
that dominates the spectrum in the optical region.  

The first calculation of the spectrum of a black hole spherically accreting in
the ISM was by \citet{Shvartsman}, and investigated by numerous authors [some
early work includes \citet{Zeldovich,Novikov,Shapiro}]. 

\begin{figure}
\plotone{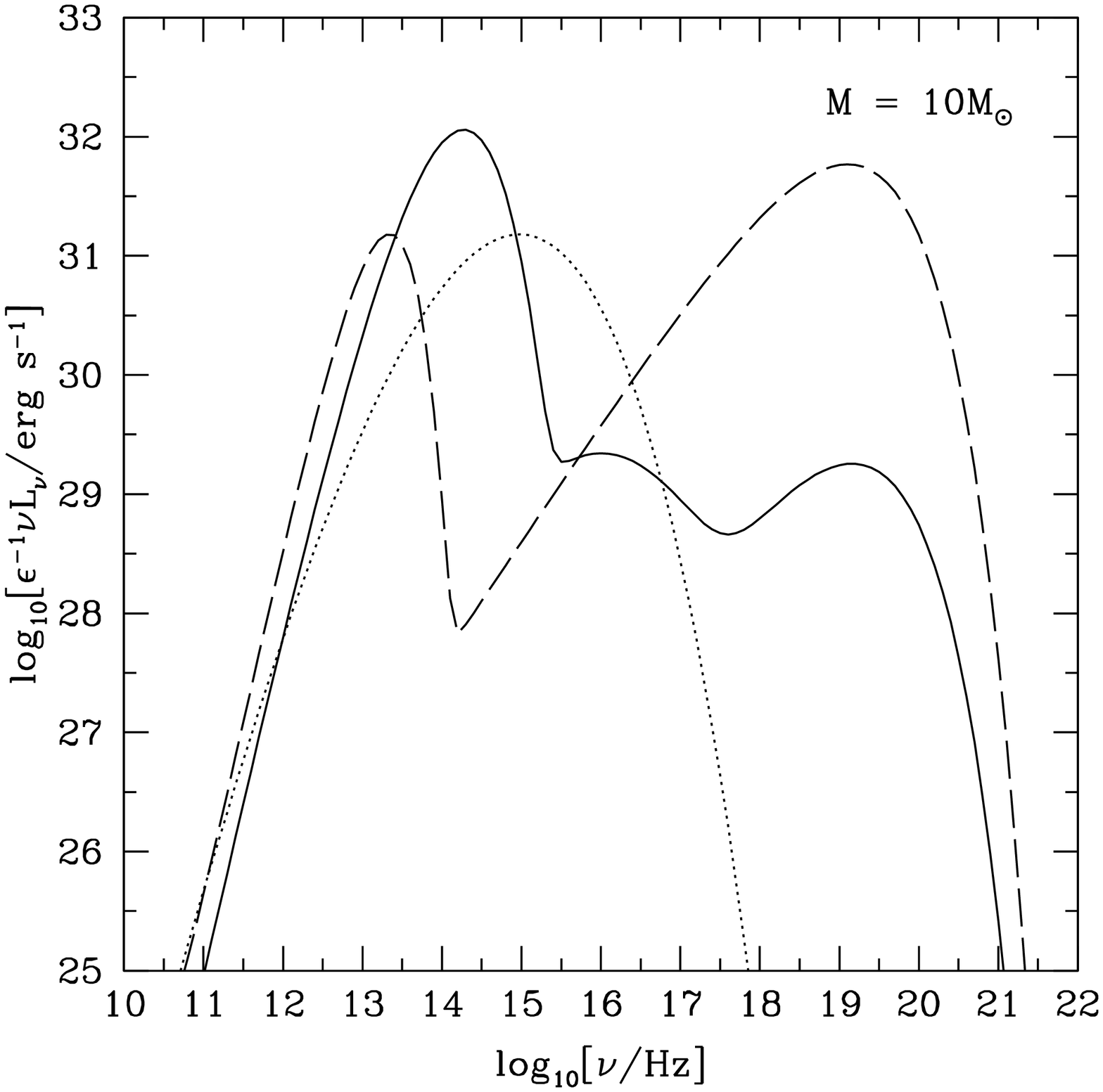}
\caption{\label{GUF} Sample models for the spectral energy distribution of a 
$10M_\odot$ black hole accreting from the ISM. The solid line is an ADAF model,
the dotted line is an IP model, and the dashed line is a CDAF model.  The
curves are meant to show the major spectral differences between the models. The
integrated luminosity of all three models is $\epsilon \dot{M}$, where
$\dot{M}=2.5\times10^{32}\ \textrm{erg s}^{-1}$ is the mass accretion rate for
a hole of mass $10M_\odot$ and velocity of $40\ \textrm{km s}^{-1}$.}
\end{figure}

The estimates for observing black holes in \citet{Heckler} used a
spherical accretion model and computed the spectrum using the method of
\citet{Ipser} (hereafter, IP).  The problem of spherical accretion was first
solved by \citet{Bondi}, and the resulting accretion flow is termed a Bondi
flow.  Spectra have been computed for such flows by \citet{Ipser} and
\citet{McDowell}.  The basic result is that interstellar magnetic
fields are drawn in and compressed in the accretion process, reaching about 10
tesla at the horizon for standard ISM conditions.  The resulting synchrotron
radiation can be quite high.  The spectral energy distribution in such a model
is indicated in Fig.\ \ref{GUF} for a $10M_\odot$ RBH.  The characteristic
spectrum is a power-law (albeit with different power laws indices above and
below the synchrotron peak).  

Recently, \citet{Igumenshchev} have shown that spherical accretion may
not be as simple as previous studies would indicate, due to the fact that the
dynamical effects of magnetic fields on the accretion flow have not properly
been taken into account.  They perform three-dimensional magnetohydrodynamic
simulations of an initially spherically symmetric system (assuming an initially
uniform magnetic field), and show that the resulting flow is convectively
unstable.  This convective flow (which they term a convection-dominated Bondi
flow, or CDBF) has a drastically decreased accretion rate compared to a Bondi
flow (approximately 9 orders of magnitude for an RBH), which would drastically
affect the emission spectrum.  At this time there are no calculations of
emission spectra from CDBFs in the literature.  However, there are enough
similarities between CDBF and CDAF models (discussed below) that one could use
computed CDAF spectra \citep{Ball} for CDBF spectra.

Even without the instability of Bondi flow, if there is sufficient
inhomogeneity in the ISM due to density or velocity gradients, the accreting
matter will have appreciable angular momentum and can form a disk
\citep{Fryxell, Taam}.  These inhomogeneities can also lead to disk reversal.
Disk accretion has been extensively studied for 
X-ray binaries where it is believed that an accretion disk powers the X-ray
emission.  There, the disk forms when the accreting gas has non-zero angular
momentum as it falls onto the compact object, such as through a solar wind or
through Roche lobe overflow.  The hydrodynamics of such a flow was first solved
by \citet{Shakura}.  Since then, there have been a number of advancements in
the field.  In particular, for systems with low accretion rates (as expected
for RBHs), one might expect the formation of an optically thin accretion
disk. 

The instabilities and inhomogeneities that lead to disk reversal also
lead to variabilities on the same timescales mentioned in \citet{Heckler},
equation 2 (reprinted here);
\begin{equation}
\Delta t \approx 10 \left(\frac{M}{M_\odot}\right) 
\left(\frac{v}{10\ \textrm{km s}^{-1}}\right)^{-3}\  \textrm{yr}.
\end{equation}

The advection-dominated accretion flow (ADAF) model
\citep{Ichimaru,Rees,NarayanYi,AbramowiczChen} has a two-temperature structure
that results in a broadband spectra from radio to gamma rays.  \citet{Fujita}
compute spectra for a RBH using an ADAF model following the prescription of
\cite{Manmoto}.  A typical ADAF spectral energy distribution is also shown in 
Fig.\ \ref{GUF}.  Like the IP spectrum, the spectrum around optical frequencies
is a power-law synchrotron spectrum.  Unlike the IP spectrum, there is
appreciable luminosity at higher frequencies.

ADAF models are expected to be convectively unstable
\citep{NarayanYi,Begelman}, and lead naturally to convection-dominated
accretion flow (CDAF) models.  In fact, \citet{Abramowicz} have argued that
recent {\it Chandra} observations of black hole and neutron star systems
support CDAF, instead of ADAF, models.  A typical CDAF spectral energy
distribution is also shown in Fig.\ \ref{GUF}.  Superficially it resembles the
ADAF and IP spectra at optical frequencies (a power-law synchrotron spectrum
with the peak possible in the optical region).  Like ADAF models it has a
high-energy peak; in fact the ratio of X-ray and $\gamma$-ray luminosity to
synchrotron luminosity is even larger in the CDAF models than in ADAF models.

Clearly, the expected spectral energy distribution of accreting RBHs is a
complicated problem and it is impossible to specify the spectrum with complete
confidence.  Nevertheless, for our purposes we can make the general assumption
that for optical frequencies the spectrum is synchrotron, possibly with the
peak frequency, $\nu_\textrm{peak}$, in the optical region.  The location of
the synchrotron peak scales with the mass of the black hole; using the ADAF
model of \citet{Manmoto}, $\nu_\textrm{peak} \propto M^{-3/8}$. The result is a
broken power-law spectrum, possibly making the transition in the visible region
of the electromagnetic spectrum from a positive slope for $\nu <
\nu_\textrm{peak}$ to a constant negative slope for $\nu >\nu_\textrm{peak}$. 
We also assume that there is an appreciable X-ray luminosity.

In addition to the accretion luminosity, it was shown in \citet{Armitage} that
there could be a substantial (equaling or exceeding the accretion luminosity)
energy emission due to the Blandford-Znajek effect \citep{Blandford} if the
black hole is rotating.  This energy is likely emitted at $\gamma$-ray
frequencies, so we do not consider this model here.

\section{ESTIMATE OF THE NUMBER OF EXPECTED SOURCES}

There are a number of uncertain parameters that enter into the estimate of the
expected number of detected sources.  The parameters can be broadly grouped
under the following categories: (i) There are parameters that describe the
properties of the ISM.  The properties of the ISM determine the accretion
rate, and hence the total luminosity of an object. (ii) There are parameters
that specify the number density of black holes as a function of their mass and
the distribution of hole velocities with respect to the ISM.  The luminosity,
as well as the spectrum (in particular, $\nu_\textrm{peak}$) depends upon the
mass of the hole.  Naturally, the number of detected sources will scale with
the normalization of the RBH mass distribution function, i.e., the overall
density of black holes. The accretion rate depends upon the velocity of the
hole.  (iii) The emission model (IP, ADAF, CDAF, etc.) determines the spectrum
of the holes.  Because of the complexity of the emission models we must
parameterize the model spectra. (iv) Finally, observational parameters such as
spectral coverage and limiting magnitude determine the number of expected
source detections.

The total luminosity of the RBH will depend on the accretion rate.  The RBH
accretion rate is a sensitive function of the black hole mass and velocity with
respect to the ISM: for an object of mass $M$ moving supersonically with
velocity $v$, $\dot{M} \propto M^2 v^{-3}$ (for fixed ISM conditions).  Thus,
in order to make predictions or place upper limits on RBH density, it is
important to consider an appropriate model for the RBH mass function and
velocity distribution.

A natural method of determining the RBH mass function and velocity distribution
would be to start from an underlying progenitor star population and use that to
infer the RBH properties.  Thus, in this work we assume a power law mass
function and Maxwellian velocity distribution, as follows.  Let $\Phi(M,v)dM
dv$ be the number of RBH's per cubic parsec in the range $(M,M+dM)$,
$(v,v+dv)$, and assume $\Phi(M,v)$ factorizes as
\begin{equation}
\label{capphi}
\Phi(M,v) = \phi_{M}(M) \phi_{v}(v),
\end{equation}
where
\begin{equation}
\label{phim}
\phi_{M}(M) = \phi_0 \left(\frac{M}{M_\odot}\right)^{-(1+x)}
\end{equation}
and
\begin{equation}
\label{phiv}
\phi_{v}(v) = \left(\frac{2}{\pi}\right)^{1/2} 
\frac{v^2}{\sigma^3} \exp\left(-\frac{v^2}{2 \sigma^2}\right) .
\end{equation}
Here, $x$ determines the power law slope of the mass function and $\sigma$ is
the velocity dispersion.  \citet{Fryer} have derived the theoretical RBH mass
function starting from a power law progenitor mass function, and their results
suggest the slope of the power law does not change when a star becomes a
RBH. On the assumption that the RBH mass distribution is not much different
from the stellar mass distribution from which it is formed, we take $x=2$. We
assume the RBHs range in mass from 3 to 100 $M_\odot$.

We will assume that the velocity dispersion of the RBH's is similar to that of
the X-ray-binary population, $\sigma=40\ \textrm{km s}^{-1}$
\citep{Hansen,White,VanParadjis}.  

The ISM parameters that enter are the sound speed ($c_{s}$) and the mass
density ($\rho_{\infty}$) of the ISM.  The relevant black hole distribution
parameters are the local mass density of black holes ($\rho_{BH}$), the
velocity dispersion of the black holes ($\sigma$), the slope of the black-hole
mass function ($x$), and the maximum and minimum black hole masses
($M_{min},M_{max}$).  The emission model parameters include an overall factor
describing the efficiency of conversion of accreting mass to luminosity in each
bandpass (the five SDSS bands $u^*,\ g^*,\ r^*,\ i^*,\ z^*$,\ X-ray, IR, etc.):
$\epsilon_\alpha \equiv L_\alpha/\dot{M}$ where $L_\alpha$ is the luminosity in
band $\alpha$ and $\dot{M}$ is the accretion rate. Note that all the model
dependence (CDAF, IP, etc.) is hidden in the efficiencies $\epsilon_\alpha$.
Finally, the observational parameters include the solid angle of sky covered
($\Omega_{SDSS}$) and the minimum detectable flux in each bandpass
($F^{min}_\alpha)$.

\citet{Fryer} suggests that black holes receive a ``kick'' when formed, 
similar to what happens when neutron stars formed, on the order of 100 km
s$^{-1}$.  \citet{Popov} have modeled the spatial distribution of accretion
luminosity from isolated accreting RBHs and neutron stars in our galaxy.  They
show the luminosity has a toroidal structure (centered and in the plane of the galaxy) with radius of about 5 to 6 kpc for
neutron stars, and about 4 to 8 kpc for black holes.  They additionally
incorporate the effect of supernova kicks, using characteristic values of 200
km s$^{-1}$ and 400 km s$^{-1}$.

The SN kick can have a profound effect on the velocity distribution, depending
on the values of the kick velocity $k$ and velocity dispersion $\sigma$.  As an
example, consider an initial stellar population (all of which will evolve to SN
and become RBHs) with a Maxwellian velocity distribution.  We then randomly
``kick'' every object in this population.  Averaging over the kick direction,
we obtain
\begin{equation}
\phi_\textrm{kicked}(v)=\left(\frac{2}{\pi}\right)^{1/2} \frac{v^2}{\sigma^3} 
\left[\left(1+\frac{k^2}{v^2}\right)I_{0}\left(-\frac{vk}{\sigma^2}\right) 
+ 2 \frac{k}{v}I_{1}\left(-\frac{vk}{\sigma^2}\right)\right] 
\exp\left[- \frac{v^2+k^2}{2 \sigma^2} \right],
\end{equation}
where $I_{0}(x)$ and $I_{1}(x)$ are modified Bessel functions.  Whereas a
purely Maxwellian velocity distribution has very few members near $v=0$
($\phi_{v}(0)= 0$), a kicked Maxwellian velocity distribution can have a
significant amount:
\begin{equation}
\phi_\textrm{kicked}(0)=\left(\frac{2}{\pi}\right)^{1/2} \frac{k^2}{\sigma^3} 
\exp\left(-\frac{k^2}{2 \sigma^2}\right).
\end{equation}  
The ``low-velocity'' population will have a larger accretion rate, and lead to 
more sources than in our ``unkicked'' model.

Now we turn to the estimate of $N_{\alpha}$, the number of RBH's observable
in the SDSS survey in each bandpass.  We start with
\begin{equation}
dN_{\alpha} = \frac{\Omega_{SDSS}}{3} \ d_{\alpha-max}^3 (M,v) \ 
\Phi (M,v) \ dM dv,
\end{equation}
where $\Omega_{SDSS}$ is the solid angle of the SDSS, $\Phi (M,v)$ is the
distribution function from the previous section, and $d_{\alpha-max}$ is
defined for each bandpass as the effective maximum distance to a detectable
source:
\begin{equation}
\label{ddef}
L_\alpha = 4 \pi F^{min}_\alpha d_{\alpha-max}^2 = \epsilon_\alpha\dot{M}.
\end{equation}

Since we parameterize the luminosity in each bandpass by a single parameter
$\epsilon_\alpha$, we must calculate $\dot{M}$.  The accretion rate is well
approximated by
\citep{Bondi}:
\begin{equation}
\dot{M} = \pi r_A^2 \rho_{\infty} \sqrt{v^2 + c_s^2}
\end{equation} 
where the accretion radius is defined as
\begin{equation}
r_A = \frac{2 G M}{v^2 + c_s^2},
\end{equation}
with $v$ the velocity, and $c_s$ the sound speed.  Let $\beta_s \equiv v/c_s$.
Then
\begin{equation}
\label{mdot}
\dot{M} = \frac{4 \pi G^2 M_\odot^2 \rho_{\infty}}{c_s^3} 
\left(\frac{M}{M_\odot}\right)^2 \left(1 + \beta_s^2 \right)^{-3/2}.
\end{equation}

Solving for $d_{\alpha-max}$ from its definition in equation (\ref{ddef}),
and using equation (\ref{mdot}) for
$\dot{M}$, we obtain
\begin{equation}
d_{\alpha-max} =  \left(\frac{L_\alpha}{4 \pi F^{min}_\alpha}\right)^{1/2} 
  = (G M_\odot) \epsilon_\alpha^{1/2} \rho_\infty^{1/2}
\left(F^{min}_\alpha\right)^{-1/2} 
c_s^{-3/2} \left(\frac{M}{M_\odot}\right) (1 + \beta_s^2)^{-3/4}.
\end{equation}

In general, one expects $\epsilon_\alpha$ to be a function of $M$, $\dot{M}$,
$v$, etc.  There are not a lot of calculations of model spectral energy
densities for the mass range and accretion rates of interest to us.  Of the
relevant calculations that do exist, there are more ADAF calculations than CDAF
calculations.  So in estimating $\epsilon_\alpha$ we will be guided by the ADAF
calculations.  \citet{Manmoto} have studied the features of ADAF spectra.
Their results are consistent with most of the energy in the optical region in
a broken-power-law spectrum.  They find the peak frequency to scale as 
\begin{equation}
\nu_\textrm{peak}(M)=10^{15}\left(\frac{M_\odot}{M}\right)^{3/8}\ \textrm{Hz} .
\end{equation}
They also find that the peak value of the spectral energy distribution to be
\begin{equation}
\left[\nu L_\nu\right]_\textrm{peak}=3\times10^{-3}\dot{M} ,
\end{equation}
independent of the mass of the hole.  The value of $\left[\nu
L_\nu\right]_\textrm{peak}$ is roughly the total integrated luminosity.

\begin{figure}
\plotone{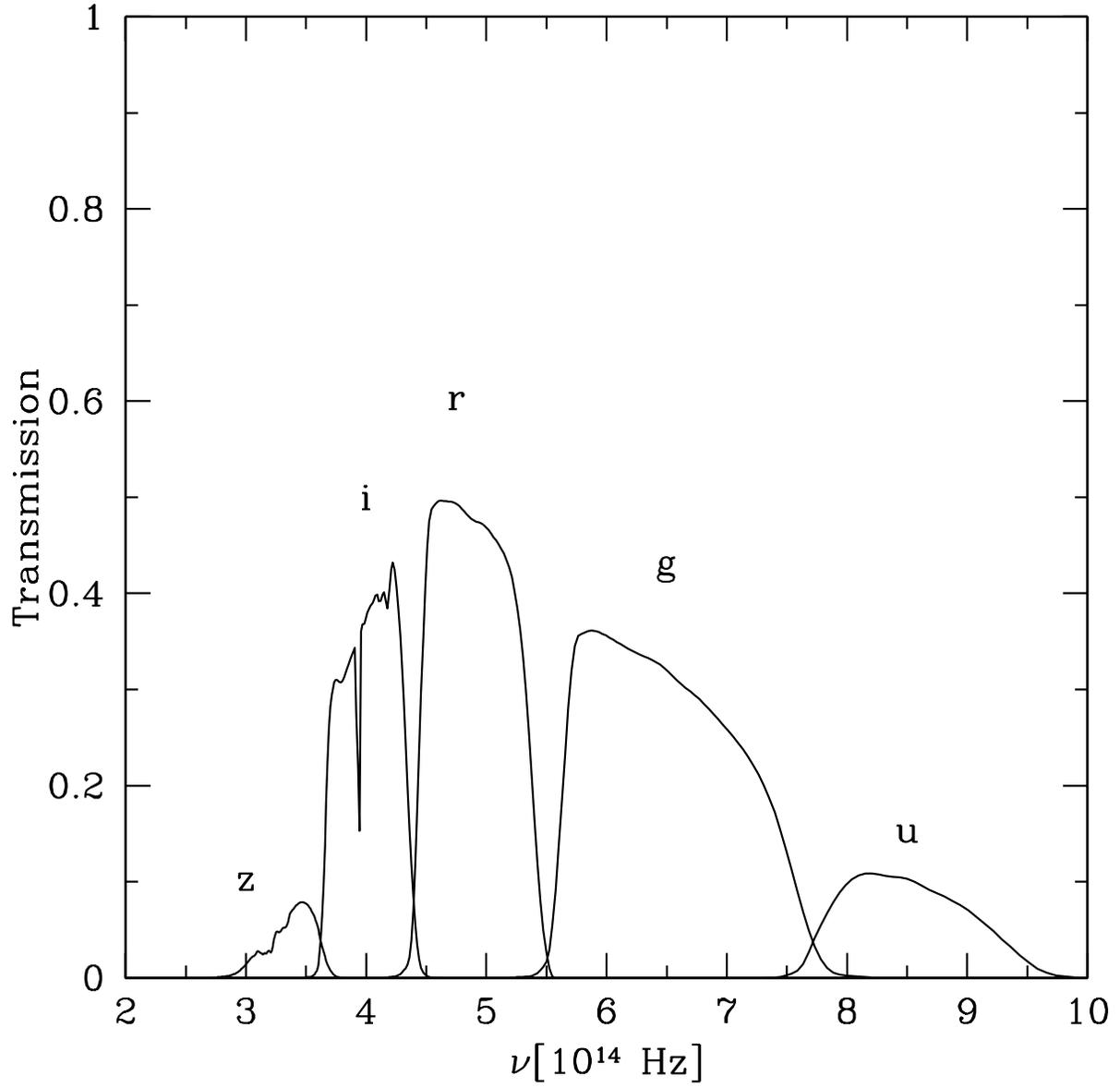}
\caption{\label{BANDPASS} The SDSS transmission curves in the five filter
bands, including atmospheric effects (airmass of 1.3), from \citet{Stoughton}.}
\end{figure}

It is convenient to express $\epsilon_\alpha$ as a product of the total
fraction of $\dot{M}$ that is radiated (integrated over all frequencies), times
the fraction radiated in band $\alpha$:
\begin{equation}
\epsilon_\alpha = \epsilon \times f_\alpha =
2\times10^{-3}\left(\frac{\epsilon}{2\times10^{-3}}\right) \ f_\alpha.
\end{equation}
The definition of $f_\alpha$ is
\begin{equation}
f_\alpha = \frac{\int_\alpha \left[\nu L_\nu\right]\left(d\nu/\nu\right)}
                {\int_0^\infty \left[\nu L_\nu\right]\left(d\nu/\nu\right)},
\end{equation}
where the ``alpha'' notation in the numerator implies integration over the
range of frequencies appropriate for band $\alpha$.  The frequencies for the
various SDSS filters are given in Table \ref{sdsstable} and Fig.\
\ref{BANDPASS}.

\begin{figure}
\plotone{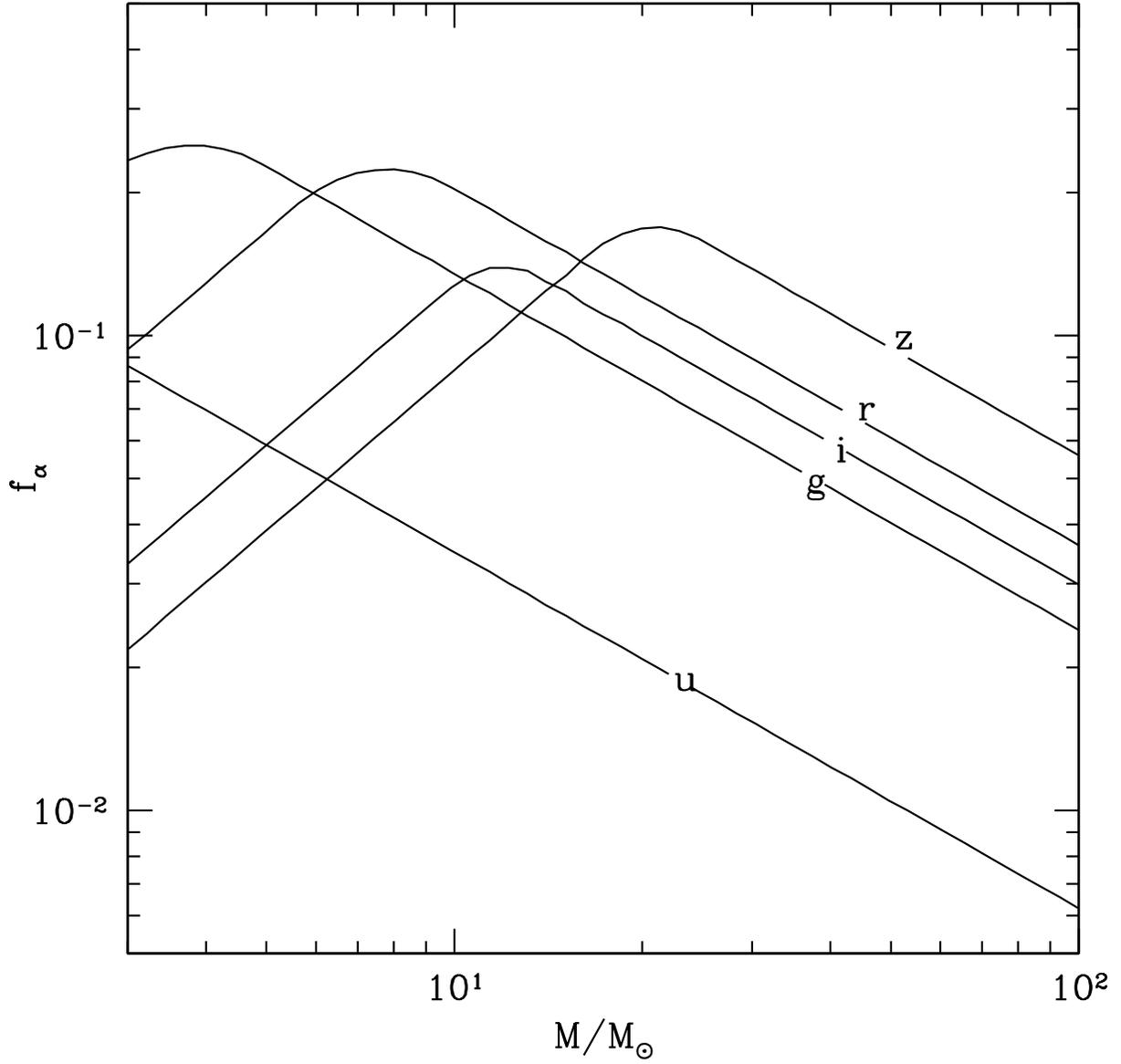}
\caption{\label{EFFICIENCY} The fraction of the luminosity in different SDSS
filters as a function of the mass of the black hole, assuming an ADAF model 
for the spectral energy distribution.}
\end{figure}

To estimate $f_\alpha$, assume the simple broken-power-law spectrum for
$\left[\nu L_\nu\right]$
\begin{equation}
\left[\nu L_\nu\right] = \left[\nu L_\nu\right]_\textrm{peak}\left\{
\begin{array}{ll}
(\nu/\nu_\textrm{peak})^3 & (\nu<\nu_\textrm{peak}) \\
(\nu/\nu_\textrm{peak})^{-2} & (\nu>\nu_\textrm{peak}) . \\
\end{array}
\right.
\end{equation}
Now using the information from \citet{Manmoto} for $\nu_\textrm{peak}(M)$ and
$\left[\nu L_\nu\right]_\textrm{peak}$, along with the information about the
SDSS filters given in Table \ref{sdsstable}, it is straightforward to calculate
$f_\alpha$ for the various SDSS filters.  The result is shown in Fig.\
\ref{EFFICIENCY}. 

For black-hole masses in the range of interest, we find $3\times10^{-1} >
f_\alpha > 6\times10^{-3}$.  While there is a mass dependence to $f_\alpha$, it
is rather complicated, and to the accuracy needed here it is adequate to assume
a constant value of $f_\alpha\sim 5\times10^{-2}$.  Therefore for
$\epsilon_\alpha$ we will use
\begin{equation}
\epsilon_\alpha=\epsilon \times f_\alpha =
10^{-4}\left(\frac{\epsilon}{2\times10^{-3}}\right) 
              \left(\frac{f_\alpha}{5\times10^{-2}}\right).
\end{equation}

Now turning to the expression for $\Phi(M,v)$ [see equations (\ref{capphi}) to
(\ref{phiv})], we must first normalize the RBH distribution.  Using equations
(\ref{phim}) and (\ref{phiv}), we find
\begin{eqnarray}
\rho_{RBH} & = & \int dM \ M \int dv \ \Phi(M,v) \nonumber \\
        & = & \phi_0 \int dM M \left(\frac{M}{M_\odot}\right)^{-(1+x)} 
                      \int dv \left(\frac{2}{\pi}\right)^{1/2} 
                \frac{v^2}{\sigma^3} \exp\left(-\frac{v^2}{2 \sigma^2}\right).
\end{eqnarray}
It is useful to define a dimensionless mass for the black hole, $\mu \equiv
M/M_\odot$.  The velocity distribution is already normalized to unity, so,
\begin{equation}
\rho_{RBH} = \phi_0 M_\odot^2 \int_{\mu_{min}}^{\mu_{max}} d\mu \ \mu^{-x} 
           = \phi_0 M_\odot^2 I(x) \equiv \rho_{DM} f,
\end{equation}
where $f$ is the fraction of the local dark matter density in RBHs.  We
use the value of $\rho_{DM}=0.01 M_\odot$/pc$^3$ from \citet{Gates}.
$I(x)$ is
of order unity; e.g., $I(x)=1/3$ for $x=2$ and $\mu_{min}=3$.  Solving for
$\phi_0$ in terms of $f$, $\rho_{DM}$, and $I(x)$ , the result is
\begin{equation}
\phi_0 = \frac{f \rho_{DM}}{M_\odot^2 I(x)} .
\end{equation}
Substituting this into $\Phi(M,v)$ and defining the ratio of the ISM sound 
speed to the black-hole velocity dispersion to be $\zeta \equiv c_s/\sigma$, 
\begin{equation}
\Phi(M,v) = \frac{4 \pi \rho_{DM} f }{ (2\pi)^{3/2} M_\odot^2 \sigma I(x) } 
\mu^{-(1+x)} \zeta^2 \beta_s^2 \exp\left(-\zeta^2 \beta_s^2 /2 \right) .
\end{equation}
With this expression for $\Phi(M,v)$, we find $dN_{\alpha}$ to be
\begin{eqnarray}
dN_{\alpha} & = & \frac{\Omega_{SDSS}}{3}(G M_\odot)^3\epsilon_\alpha^{3/2}
\rho_\infty^{3/2} \left(F^{min}_\alpha\right)^{-3/2} c_s^{-9/2} 
\mu^3 (1 + \beta_s)^{-9/4} \nonumber \\
& & \times \frac{4 \pi \rho_{DM} f }{ (2\pi)^{3/2} M_\odot^2 \sigma I(x) } 
\mu^{-(1+x)}\zeta^2\beta_s^2\exp\left(-\zeta^2\beta_s^2 /2\right)d\mu\ d\beta_s
\nonumber \\
    & = & \textrm{const} \times \frac{\zeta^3}{I(x)}\mu^{2-x} \beta_s^2 
(1 + \beta_s)^{-9/4} \exp\left(-\zeta^2 \beta_s^2 /2 \right) d\mu \ d\beta_s ,
\end{eqnarray}
where the constant in the above expression is given by
\begin{equation}
\textrm{const} = \frac{\Omega_{SDSS}}{3} (G M_\odot)^3 \epsilon_\alpha^{3/2} 
\rho_\infty^{3/2} \left(F^{min}_\alpha\right)^{-3/2} c_s^{-9/2} 
\frac{4 \pi \rho_{DM} f } { (2\pi)^{3/2} M_\odot } .
\end{equation}
Integrating over $\mu$ and $\beta_s$, 
\begin{eqnarray}
N_{\alpha} & = & \textrm{const}\times \frac{\zeta^3}{I(x)} 
\int_{\mu_{min}}^{\mu_{max}} 
d\mu\  \mu^{2-x} \int_{0}^{\infty} d\beta_s \beta_s^2 (1 + \beta_s)^{-9/4}
\exp\left(-\zeta^2 \beta_s^2 /2 \right).
\end{eqnarray}
With the final definitions 
\begin{eqnarray}
I'(x) & = & \int_{\mu_{min}}^{\mu_{max}} d\mu \ \mu^{2-x} \nonumber \\
J(\zeta) & = & \int_{0}^{\infty} d\beta_s \beta_s^2 (1 + \beta_s)^{-9/4}\exp
\left(-\zeta^2 \beta_s^2 /2 \right) ,
\end{eqnarray}
we obtain
\begin{eqnarray}
N_{\alpha} & = & \frac{\Omega_{SDSS}}{3}(G M_\odot)^3\epsilon_\alpha^{3/2} 
\rho_\infty^{3/2} \left(F^{min}_\alpha\right)^{-3/2} c_s^{-9/2} 
\frac{4 \pi \rho_{DM} f }{ (2\pi)^{3/2} M_\odot } 
\left(\frac{I'(x)}{I(x)}\right) \zeta^3 J(\zeta) .
\end{eqnarray}
Using $x=2$, $\mu_{min} = 3$, $\mu_{max} = 100$, $I'(x)/I(x) = 300$, 
and using $\zeta = 16.6/40 \approx 0.415$, $\zeta^3 J(\zeta) \approx 5.82 
\times 10^{-2}$, we find
\begin{eqnarray}
\label{nobs}
N_{\alpha} & = & 10^6 f 
\left(\frac{\epsilon}{2\times10^{-3}}\right)^{3/2} 
              \left(\frac{f_\alpha}{5\times10^{-2}}\right)^{3/2}
\left(\frac{F^{min}_\alpha}
          {10^{-15}\ \textrm{erg cm}^{-2}\,\textrm{s}^{-1}}\right)^{-3/2}
 \nonumber \\ 
& & 
\times \left(\frac{c_s}{16.6\ \textrm{km s}^{-1}}\right)^{-3/2}
\left(\frac{\rho_{\infty}}{10^{-24}\ \textrm{g cm}^{-3}}\right)^{3/2} 
\left(\frac{\sigma}{40\ \textrm{km s}^{-1}}\right)^{-3/2} .
\end{eqnarray}

This analysis does not account for interstellar reddening.  The limiting
magnitudes corresponding to $F^{min}_\alpha=10^{-15}\ \textrm{erg
cm}^{-2}\,\textrm{s}^{-1}$ depends of course on the bandpass and the source
spectrum, and may be inferred from Table \ref{sdsstable}.  For illustration, in
the $r^*$ band, a flux of $10^{-15}\ \textrm{erg cm}^{-2}\,\textrm{s}^{-1}$
corresponds roughly to $m=23$.

Since the local ISM is not homogeneous, we may ask to what distance one
could detect a RBH of mass $M$ moving at velocity $v$.  It is
\begin{eqnarray}
d_{\alpha-max}&=&200\,\textrm{pc}
\left(\frac{\epsilon}{2\times10^{-3}}\right)^{1/2} 
              \left(\frac{f_\alpha}{5\times10^{-2}}\right)^{1/2}
\left(\frac{F^{min}_\alpha}
           {10^{-15}\ \textrm{erg cm}^{-2}\,\textrm{s}^{-1}}\right)^{-1/2}
\nonumber \\
& & \times \left(\frac{c_s}{16.6\ \textrm{km s}^{-1}}\right)^{-3/2} 
\left(\frac{\rho_{\infty}}{10^{-24}\ \textrm{g cm}^{-3}}\right)^{1/2} 
\left(\frac{M}{M_\odot}\right) 
\left[1 + \left(\frac{v}{c_s}\right)^2 \right]^{-3/4} .
\label{eq:Dist}\end{eqnarray}

It should be emphasized that the expressions for $N_{\alpha}$ and
$d_{\alpha-max}$ contain parameters that can feasibly range over a few orders of
magnitude due to variations in the ISM ($\rho_\infty \sim 10^{-22} \textrm{ to } 10^{-25} \textrm{ g cm}^{-2}, c_s \sim 0.5 \textrm{ to } 50 \textrm{ km s}^{-1}$) 
or spectral model ($f_\alpha \sim 3\times10^{-1} \textrm{ to } 6\times10^{-3},\epsilon \lesssim 10^{-2}$).   

Note that we are not including any spatial dependence in 
either $\Phi(M,v)$ or $\sigma$.  Assume that the RBH population has a disk scale 
height $H\sim 270-590$ pc \citep{Agol}.  The typical distance $Z$ above the 
galactic disk is $d_{\alpha-max}/3$, giving a $Z \sim 400$ pc for a
6 $M_\odot$ RBH (the mean mass of the population).  By normalizing the RBH density to the local halo dark matter
density, not incorporating the exponential fall-off in spatial density implies
we are overestimating the number of candidates, but only by a factor no 
greater than roughly $e$.  Given the above noted uncertainties in the other
model parameters, this is not of large concern.

The possibility of increasing the effective distance by assuming the RBH is in
a molecular cloud has been discussed by \citet{Grindlay} and by
\citet{Campana}.  The increased accretion rate inside a molecular cloud could
raise the peak luminosity by a factor of ten, thereby increasing the maximum
distance of detection by a factor of three.  However, the filling factor of
molecular clouds is only about $1\%$, so it is unclear whether these sites
offer the best possibility for detection.

\section{SEARCH STRATEGY}

\subsection{Optical detection in the SDSS Early Data Release}

The SDSS consists of both a photometric survey and spectroscopic follow up of
selected targets.  Our strategy for finding RBH's involves searching first
through the photometric data with some selection criteria.  

The SDSS has five filter bands, denoted $u,\, g,\, r,\, i,\, z$, and
images are taken in every band.  
The filter transmission
curves are given in Fig.\ \ref{BANDPASS}, and further information is given in
Table \ref{sdsstable}.  Thus, some knowledge of the spectrum can be extracted
from the photometric data (which is a five bin spectrum).  This is done by
looking at the SDSS colors of objects.  For this analysis, we utilize
the four standard differences between adjacent bands, and make selections 
using this four dimensional color space.

For RBH's, the dominant source of emission in the SDSS optical bands is due to
synchrotron emission.  As discussed in \S 2, we can approximate the spectrum
using a broken power law.  For $\nu < \nu_\textrm{peak}$, the slope is
Rayleigh-Jeans ($\nu L_\nu \propto \nu^{3}$).  This would result in a ``blue''
spectrum.  For $\nu > \nu_\textrm{peak}$, the slope depends on the electron
energy distribution and approaches a constant negative value (we assume $\nu
L_\nu \propto \nu^{-2}$).  This would be a ``red'' spectrum.

\begin{figure}
\plotone{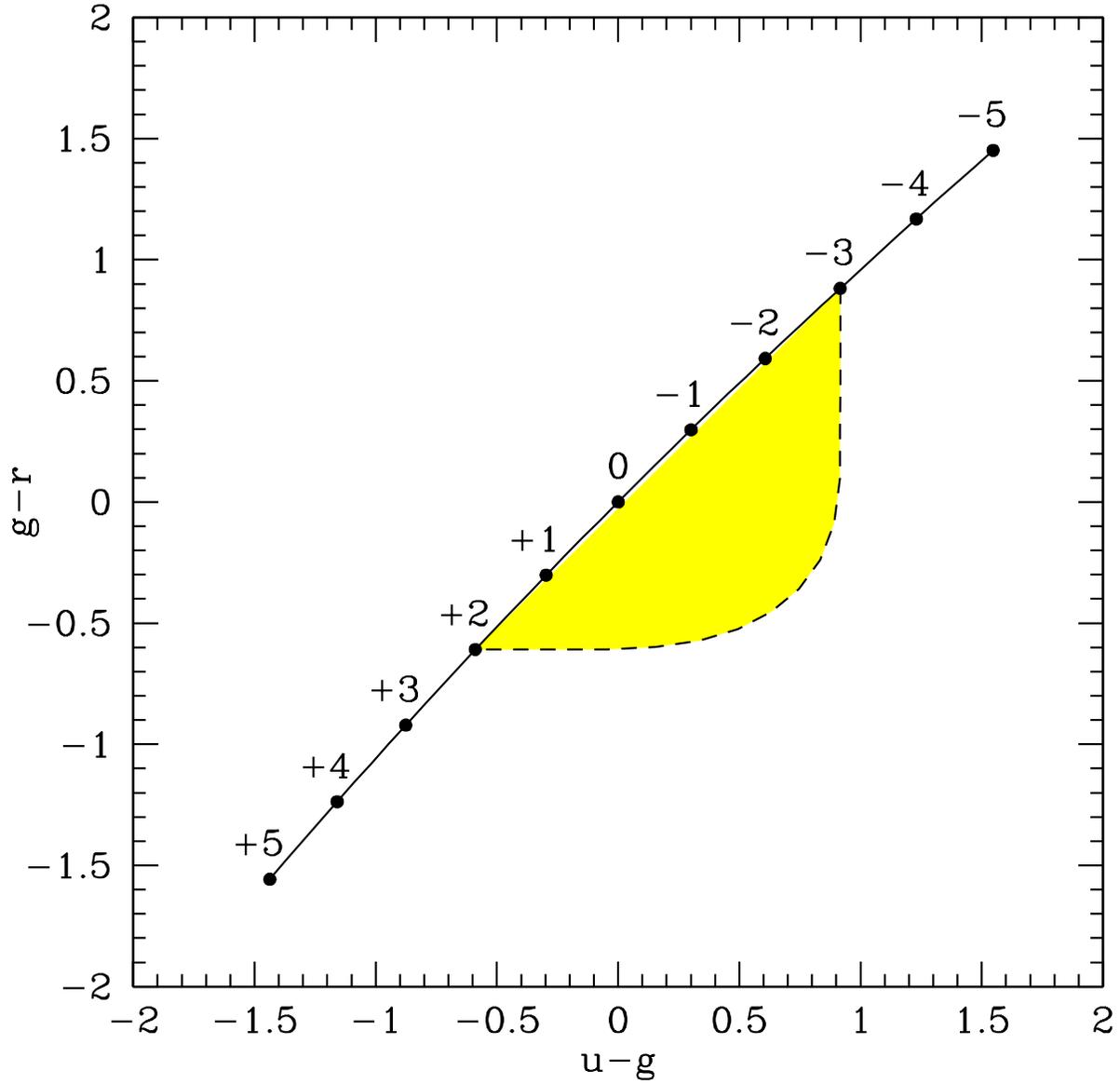}
\caption{\label{TRACKS} The tracks of a pure power law (solid line) and broken 
power law (asterisks) synthetic spectra.  The numbers indicate the power law
slope at that point on the curve.  We expect the color of RBH's to fall
somewhere in the shaded triangular region above. }
\end{figure}

We then fold this synthetic spectrum through the transmission curves to get
synthetic colors.  A pure single power-law spectrum would correspond to a point
in color--color space.  Varying the power law index traces out an approximately
straight line in color--color space as illustrated in Fig.\ \ref{TRACKS}.  It
is quite reasonable that the spectrum would be a broken power law.  If the
spectrum is approximated as a broken power law, with the break somewhere in the
SDSS sensitivity region, then in the color--color diagram the source would
appear somewhere on the dashed curve of Fig.\ \ref{TRACKS}, with its location
determined by the exact location of $\nu_\textrm{peak}$.  Of course the curve
is terminated on the line corresponding to the power-law slopes in the limiting
regimes.  In reality, one does not expect a sharp transition between the two
power laws.  Rather it is more reasonable to assume some smooth transition
between the limiting power-law slopes.  As the spectrum becomes ``flatter''
(same power law across adjacent bands), the object approaches the straight line
in color-color space.  Thus, any transitional behavior will fall within the
triangular region defined by the single power law and broken power law curves
in the color-color diagrams.  This region is indicated on Fig.\ \ref{TRACKS}.
For the four linearly independent colors, we can make six independent
color-color diagrams (i.e., six projections from the four-dimensional
multi-color space), each with a defined triangular region.

\begin{figure}
\plotone{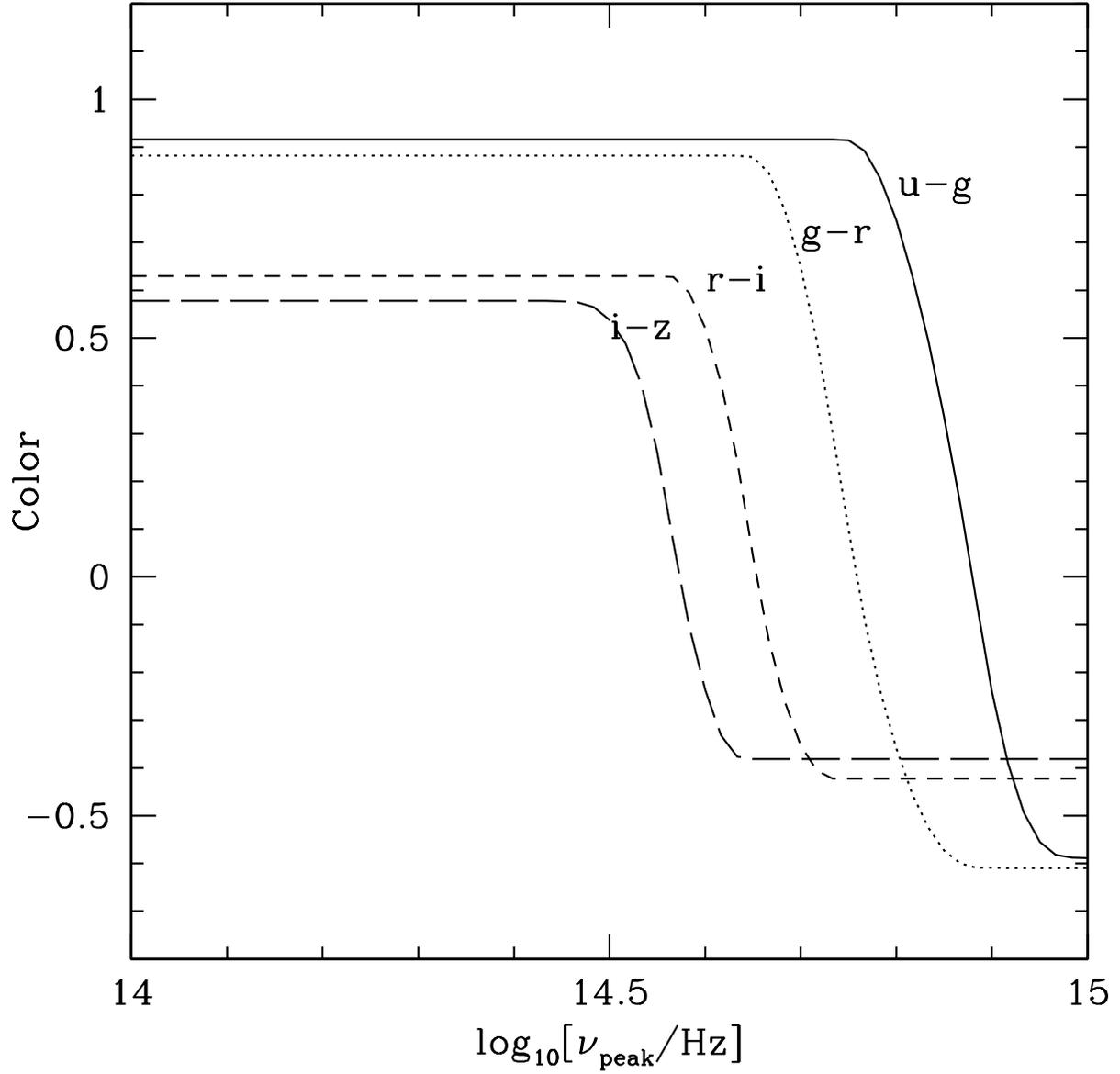}
\caption{\label{COLORTRACKS} This figure shows how the SDSS colors change with 
peak frequency.  The assumed spectrum is a
broken power law, changing from $+2$ to $-3$.  }
\end{figure}

As mentioned, the peak of the spectrum depends on the RBH mass.  Using the ADAF
peak-frequency--mass scaling mentioned above, different mass holes show up in
different regions of the color--color diagrams.  The six two-color diagrams may
have differing degrees of usefulness for our purposes.  See Fig.\
\ref{COLORTRACKS} for a plot of each color vs.\ black hole mass.  High RBH mass
corresponds to low peak frequency, thus it is the redder colors ($r-i$,
$i-z$) which turn over before the bluer colors ($u-g$, $g-r$) as we
decrease RBH mass.  

Even with the above ``cuts'' there will be many ``normal'' objects within the
triangular region since the SDSS will detect about 100 million objects on the
sky.  See Fig.\ \ref{STARS} for a selection of objects taken from the
photometric survey so far.  The large swath of objects extending up and to the
right is the stellar locus.

The strategy of searching the SDSS database for objects which fall within the
triangular region in each of the six color-color diagrams is only the first
step.  So that we will not be overloaded with background objects (normal stars,
QSOs, etc.), the next step is to correlate objects within the triangular
region with the ROSAT X-ray catalog.

\subsection{X-ray detection}

Many detections of black hole binaries have been in the X-ray regime, and
it is from this data that some of their properties can be determined.  As noted
in \citet{Fujita} and \citet{Agol}, an optimum search strategy for finding RBH's might involve
first looking in the X-ray, as some spectral models predict that the X-ray
emission would be more easily detected there than in the optical.

The ROSAT All Sky Survey (RASS) \citep{Voges} was performed with the ROSAT
X-ray satellite shortly after it was launched in 1990.  It covers the energy
band 0.1 to 2.4 keV and is the most sensitive all-sky X-ray survey to date.
The SDSS has cross-listed the RASS Bright-Source (BSC) and Faint-Source 
Catalogs (FSC) in its database, so that it is possible to
quickly determine whether or not a SDSS source is a bright X-ray source.

To be included in the RASS BSC, a source must have at least 15 source photons.
For typical exposure times, this translates into a limiting photon count rate
of 50 counts ksec$^{-1}$ \citep{Voges}.  This can be converted into a flux
limit by assuming a spectral model.  Assuming a power law spectrum with slope
ranging from $-1$ to $-3$, the flux limit is (1.9 to 5.2) $\times$ 10$^{-13}\
\textrm{erg cm}^{-2}\,\textrm{s}^{-1}$.  To be included in the RASS FSC, a
source must have at least 6 source photons.  The corresponding limiting photon
count rate and flux limit are assumed to scale with the number of source
photons (6/15), giving 20 counts ksec$^{-1}$ and (0.8 to 2.1) $\times$
10$^{-13}\ \textrm{erg cm}^{-2}\,\textrm{s}^{-1}$, respectively.  Due to
differences in exposure time in different areas of the sky, the count rate for
detected sources may fall below the above listed limiting count rates.  

The typical position error for a RASS source is 30''.  We identify a SDSS object
with a RASS source if the difference in position is less than the RASS position
error for that source.  Thus it is possible to have more than one SDSS object
identified with a single RASS source.  This implies that some of our matches may
be accidental, but given the size of the RASS error circle, this can only be 
resolved with higher spatial resolution follow-up X-ray observations.

We can investigate the possibility of ``accidental'' identifications as 
follows.  The RASS has 124,730 objects (both the BSC and FSC) distributed over
essentially the entire sky, giving an average object density of 3.02 obj/deg$^2$.
The EDR has about 2.1 million photometric objects, over the 462 deg$^2$ the EDR 
contains this gives an average object density of 4600 obj/deg$^2$.  Taking just
those EDR objects identified with RASS sources (EDR objects falling within the error
radii of RASS objects and listed in the ROSAT sub-catalog of the EDR); there are
38,404 objects, giving an average object density of 83.1 obj/deg$^2$.  

Just comparing the average object density of RASS sources (3.02 obj/deg$^2$) to that
of EDR objects identified with those RASS sources (83.1 obj/deg$^2$), we see that if a 
one to one correspondence of an EDR object with a RASS source is expected, then ``accidental''
identifications must outnumber ``real'' identifications by (on average) 83.1/3.02 $\approx$ 28:1.
This ratio obviously will vary across the sky; taking only those RASS sources
that fall within the EDR survey area (rather than taking an all sky average), the proportion becomes 22:1.

Thus, while not every EDR objects has a RASS object match (only 83.1/4600 $\approx$ 2\% do),
a RASS object has, on average, 22 EDR matches by virtue of the large error radius (again, this will vary for 
any given individual RASS source).  This would seem to imply that the criterion of requiring a RASS source
will yield no new information about the candidates.  However, due to the large RASS error circle, it is not 
possible to say that a specific EDR object identified with a RASS source does not have X-ray activity, as a
RASS source may have contributions from more than one better resolved objects.  In effect, there can be no
``accidental'' identifications without follow-up X-ray observations, since a 1:1 correspondence cannot in 
general be expected. So, while the presence of a RASS object match doesn't definitively prove that a specific EDR 
object has X-ray emission, the chances of it being an X-ray emitter are better than an EDR object with no RASS match.

\section{RESULTS}

The Science Archive Query Tool (known as sdssQT) was programmed to search the
SDSS Early Data Release (EDR) database.  The EDR database covers 462 square
degrees on the sky (about 5\% of the entire survey volume) and contains 3.7
million photometrically detected objects.  Only objects considered to be
``Primary Survey Objects'' were considered in this search; there are 2.1
million primary survey objects in the EDR.  Since we are looking for point-like
objects, we use the Point-Spread Function (PSF) magnitudes with reddening
corrections.

To be considered a candidate, the object must: 1) be a ``good'' point-like 
photometric 
object (omit extended objects and objects flagged as BRIGHT, EDGE, BLENDED or 
SATUR) with PSF
magnitude errors less than 0.20, 2) have PSF colors that fall within the 
triangular
regions in color--color space (after reddening corrections), and 3) have a 
RASS detection ({\it i.e.} be within the error box of a RASS source).  These
cuts returned 87 objects (out of an initial database of 2.1 million objects).
The results are given in Table \ref{rocky1-47}.

Some of these objects have been observed by the SDSS spectroscopically (see \citet{Stoughton} and
references therein for information on the spectroscopic survey) as well
and identified on the bases of those spectra.  Of the 87 objects, 32 have been
identified spectroscopically as 26 QSOs and 6 stars.  The latter include 1 CV, 
1 WD and 4 others tentatively identified as F stars.

\begin{figure}
\plotone{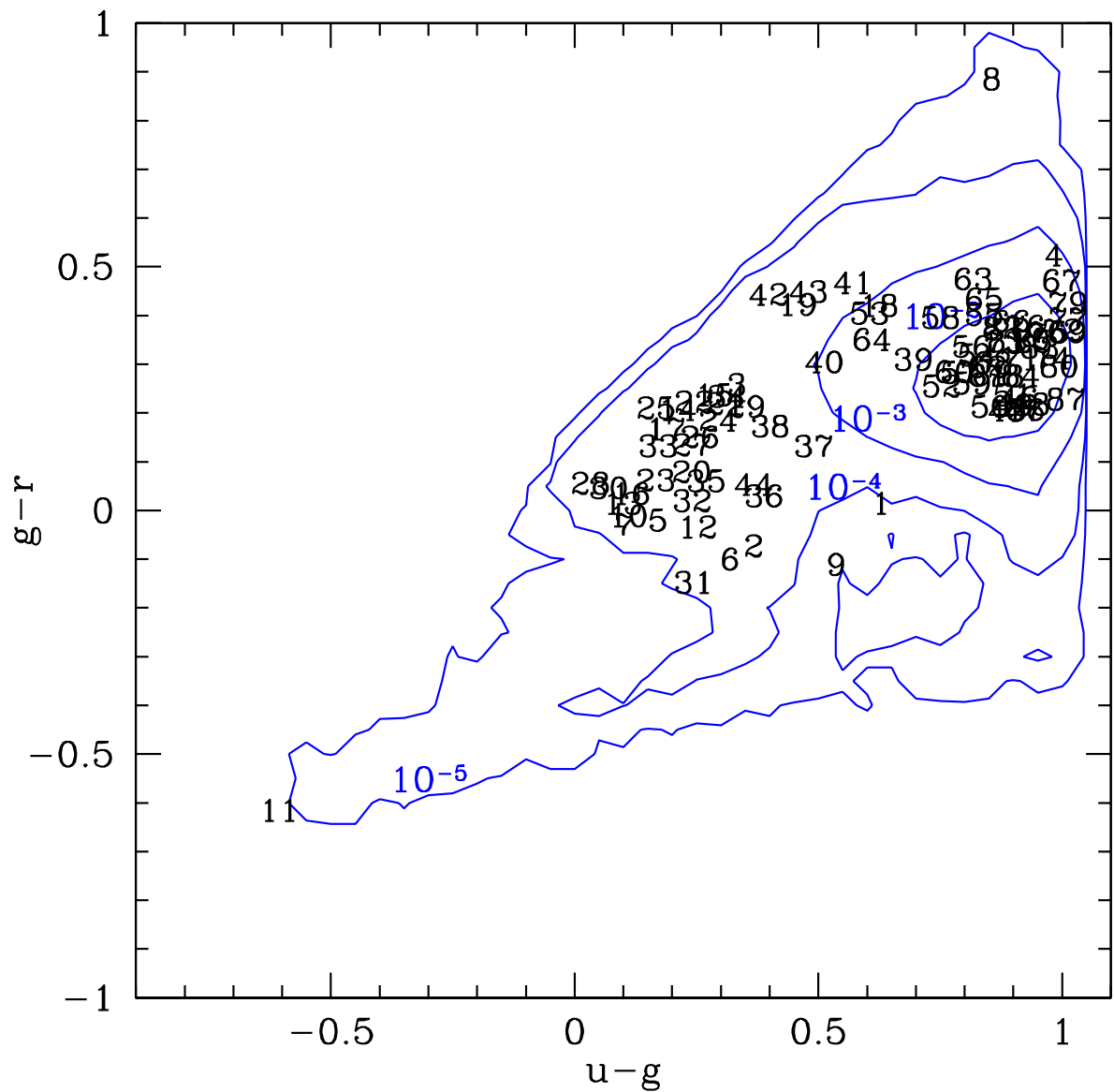}
\caption{\label{CONTOUR} Colors $u-g$ vs.\ $g-r$ of the 
87 Primary SDSS Early Data Release and RASS detected objects.  Contours show
the density of objects in our color-color diagram area that do not have RASS
detections (i.e., the background object locus).  Candidate objects are plotted
by their ordering from 
Table\ \ref{rocky1-47}.  Magnitudes are PSF and reddening
corrected.}
\end{figure}

Relaxing the third criterion (RASS detection), about 150,000 objects were
returned.  A number density of these objects was
computed and plotted as a contour with logarithmically spaced density contours.
See Fig.\ \ref{CONTOUR} for plots of the candidate objects against the 
background object density contours.  

Of those 150,000 objects that fall within our color selection region, 2939 have
been targeted spectroscopically.  As noted in \citet{Heckler}, it is possible for
RBHs to fall in the QSO selection region and have spectra taken of them.  Since they
are not quasars (or any other type of known object), they would not be identified
on the basis of these spectra and would be classified as unknown (in the SDSS database).  
In the 2939 that
have spectra, 1877 have been classified as stars, 37 as galaxies, 996 as QSOs, 
and 29 as unknown.  We list those 29 ``unknown'' objects in Table\ \ref{spectrocandidates}. 
Six of those 29 objects were later identifed as 5 QSOs and 1 WD.  Note that
object \#18 in Table\ \ref{spectrocandidates} is the same as object \#3 in Table\ \ref{rocky1-47}.

\begin{figure}
\plotone{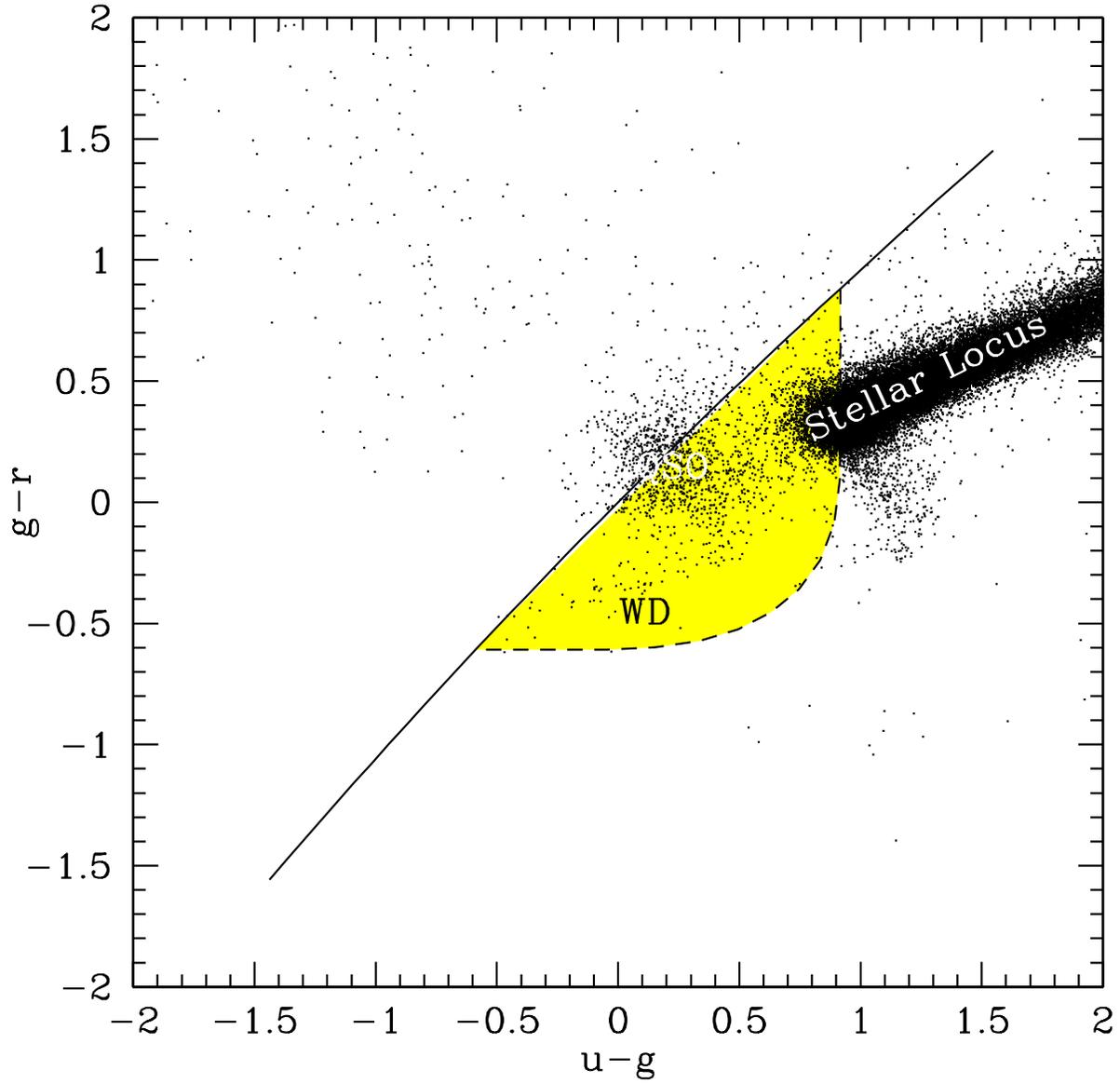}
\caption{\label{STARS} A sample of 43708 ``stars'' from the Early Data Release 
in a 10$^{o}$ slice in RA.  Noted are the positions of the stellar, QSO and
WD loci.}
\end{figure}

As seen in Fig.\ \ref{STARS}, our selection region intersect
substantially with the loci of ``normal'' stars, white dwarves and QSOs (these
we refer to as background objects).  It
is therefore not surprising that we will obtain these objects in our search,
considering both that they themselves can be X-ray active and the possibility
of ``accidental'' identification with a RASS source.
Indeed, from Fig.\ \ref{CONTOUR} it appears that most of our candidates fall 
on the background loci.

In order to find those objects in our candidate list which are less likely to be
background (star, QSO, etc.), we compute the relative overdensity $\eta$ of our candidate
objects in the 4D color space.  The 4D space is first split up into hypercubic bins of side
length 0.25 magnitudes.  The 87 candidate (RASS cut) objects and 156,563 (no RASS cut)
background objects are then histogrammed.  For every bin in which there is a candidate object, the
overdensity for that object is given by
\begin{equation}
\label{overdensity}
\eta = \left(\frac{\textrm{\# candidate objects in bin}}{\textrm{\# background objects in bin}}\right)\left(\frac{156,563}{87}\right)
\end{equation}
Note that since every candidate object also appears as a background object, this overdensity has a
maximum value of 156,563/87 $\sim$ 1800. 

\begin{figure}
\plotone{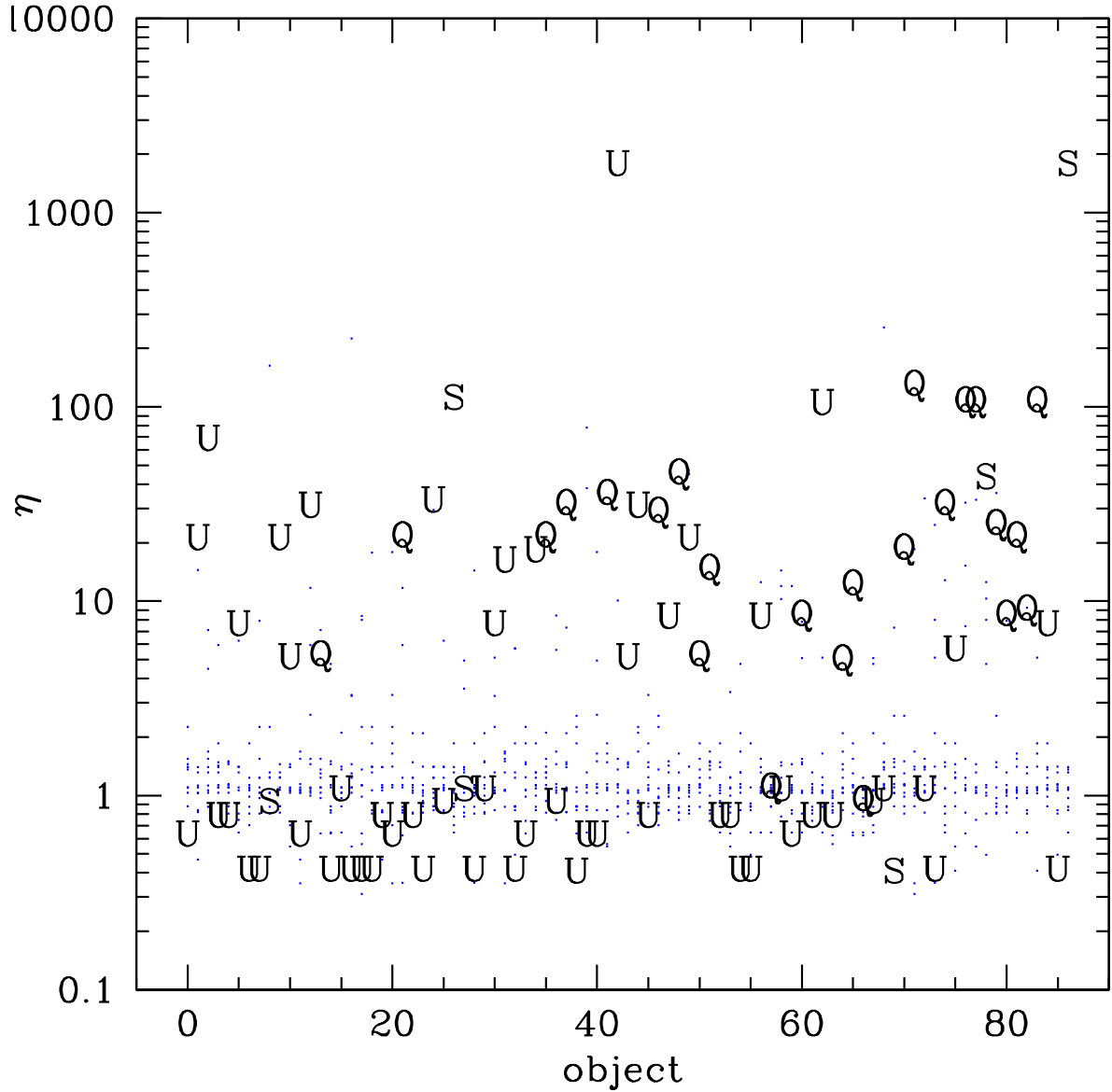}
\caption{\label{bgratioplot} The candidate object overdensity
$\eta$; small blue dots are overdensities from 10 random samples of 87 
objects from the background (none of these 870 random objects are necessarily
our candidates).  The 87 candidates are plotted (in no particular order) by 
spectral type,
if known: {\bf U}nknown (not targeted), {\bf S}tar, or {\bf Q}SO.}
\end{figure}

Using the 
synthetic object colors from \citet{Fan} (and as noted earlier, from Fig.\ \ref{STARS}),
we see that our color selection will pick up (at least) ``normal'' stars, QSOs of 
$z \lesssim 3$, WDs and Compact Emission Line Galaxies (CELGs).  Given this fact, and
the large ``accidental'' RASS identification rate computed in the last section, we might
have expected the RASS selection to effectively be a random sampling of the background  
objects.  To demonstrate this, 10 realizations of a random sampling of 87 objects from the
background (of 156,563) objects was performed, and the subsequent values of $\eta$ are
plotting in Fig.\ \ref{bgratioplot}.  As expected, most fall around $\eta \approx 1$ with
some upwards scatter.  Also plotted are the values of $\eta$ for our 87 candidates.  
Given the difference between the two populations, this implies that the RASS selection
is not dominated by ``accidental'' identifications.  In fact, note that QSOs appear
to be preferentially selected by requiring a RASS detection, as RBHs hopefully will be
as well.  This alleviates, to some extent, some worry about our 
candidate list being dominated by ``contaminant'' objects.  

We may, however, further attempt to rule out QSOs and CELGs 
as contaminants by looking for proper motion among our candidate objects.  
Being extragalactic objects, QSOs and CELGs will have no detectable proper 
motion.  

The proper motion $\pi$ is given by
\begin{equation}
\label{propermotion}
\pi \approx \frac{v}{d} \sin \theta
\end{equation}

\noindent where $\theta$ is the angle between the RBH velocity $v$ and the line of
sight.  To compute typical proper motions, we first need typical velocities
and distances;
\begin{equation}
\bar v = \int v \phi_v(v) dv = 1.6 \sigma
\end{equation}

To compute a typical distance, we first compute the number density of RBHs.
\begin{eqnarray}
n_{RBH} & = & \int \phi_M(M) dM \nonumber \\ 
        & = & \frac{f \rho_{DM}}{M_\odot I(x)} \int_{\mu_{min}}^{\mu_{max}} d\mu \mu^{-(1+x)} \nonumber\\
        & = & f \left(\frac{\rho_{DM}}{M_\odot}\right) \frac{I''(x)}{I(x)} \nonumber \\
        & = & 5.15\times10^{-4} \textrm{ pc}^{-3} \left(\frac{f}{0.3}\right)\left(\frac{\rho_{DM}}{0.01 M_\odot \textrm{ pc}^{-3}}\right)
\end{eqnarray}

Compare this with the value $8.0\times10^{-4}$ pc$^{-3}$ computed by 
\citet{Shapiro2} using a Salpeter initial mass function.

Taking a typical separation distance $d_{sep} = n_{RBH}^{-1/3}$,
\begin{equation}
d_{sep} = 12.5 \textrm{ pc} \left(\frac{f}{0.3}\right)^{-1/3}\left(\frac{\rho_{DM}}{0.01 M_\odot \textrm{ pc}^{-3}}\right)^{-1/3}
\end{equation}

Putting this into equation (\ref{propermotion}), 
\begin{equation}
\pi = (1075 \textrm{ mas yr}^{-1}) \sin \theta \left(\frac{\bar v}{1.6 \sigma}\right)\left(\frac{\sigma}{40 \textrm{ km s}^{-1}}\right) \left(\frac{d}{12.5 \textrm{ pc}}\right)^{-1}
\end{equation}

Taking an RBH at the limit of detection (at a distance from equation (\ref{eq:Dist})), 
\begin{equation}
\pi = (17.4 \textrm{ mas yr}^{-1}) \sin \theta \left(\frac{M}{M_\odot}\right)^{-1} \left(\frac{v}{c_s}\right) \left[1 + \left(\frac{v}{c_s}\right)^2\right]^{3/4}
\end{equation}

The proper motion for this latter case rises more than linearly in velocity
because in order to be detected, higher velocity RBHs need to be closer to us
than lower velocity RBHs.  

Given these estimates, we can search the USNO-A2.0 survey \citep{Monet}, which is 
also cross-listed with the SDSS EDR.  The USNO survey was obtained from digitizing
the photographic plates of the Palomar Optical Sky Survey (from 1953), and thus
provides an opportunity to measure proper motion.  Objects from the SDSS and USNO catalog
are identified in the EDR database solely by requiring $\delta < $ 30'', where $\delta$ is the
positional difference between the SDSS and USNO object.  As the 
USNO data do not go as faint as the SDSS, there will be accidental matches to faint SDSS
objects.  Further, due to astronometric inaccuracies in the USNO survey, only matches with
$\delta >$ 1'' are included as candidates for proper motion (Hugh Harris, private communication). 
Note that these bounds on $\delta \in (1, 30)''$ provide a corresponding bound on the proper motion
$\pi \in (23, 683)$ mas yr$^{-1}$.  Accidental matches can further be reduced by comparing the blue and
red magnitudes from the USNO object to the SDSS $g$ and $r$ magnitudes and requiring 
that they not be too disparate.

Table\ \ref{propermotions} lists those objects from the X-ray selected and
spectroscopically selected candidates which may have proper motion.

\section{Conclusions}

Our basic result can be found in Tables \ref{rocky1-47} and \ref{spectrocandidates}: the 55 X-ray selected 
and 18 spectroscopically selected candidates for
nearby, isolated, accreting black holes.  The X-ray selected candidates have
been ranked by $\eta$, the object overdensity in color-color space to identify
those objects most likely to be RBHs.  Given the number of candidates we 
have obtained, we return to equation (\ref{nobs})
to see if this number is reasonable.

The factor of $10^6f$ in equation (\ref{nobs}) assumes a solid-angle
coverage of $\Omega_{SDSS}=\pi$.  The Early Data Release covers $462\
\textrm{deg}^2$, or about 4.5\% of $\pi$ steradians, and the relevant
coefficient is $4.5\times10^4f$ instead of $10^6f$.

Note that this value ($4.5\times10^4f$) of represents the total number of
RBH's observed within the SDSS EDR, not those expected to be found with
our search strategy.  Since we require a RASS detection to be considered
a candidate, the expected number is the smaller of $N_{SDSS}$ and $N_{RASS}$ 
for the sky coverage of the EDR.  Using equation \ref{nobs}, we can compute
the expected $N_{RASS}$ as follows.

Using an ADAF spectral model (which the value of $N_{SDSS}$ in equation \ref{nobs} assumes), 
$f_{RASS} \approx 10^{-2}f_{SDSS} \approx 5\times10^{-4}$.
Using $F_{RASS}^{min} = 10^{-13}$ erg cm$^{-2}$ s$^{-1}$, this gives $4f$ RBH's 
over the entire sky or $0.045f$ within the EDR.  This value of $F_{RASS}^{min}$ is
actually an overestimate by about an order of magnitude, given observed sources with
count rates below the limit.  In this case, $N_{RASS}$ within the EDR rises by a factor
of 10$^{3/2}$ to $1.4f$.

Using a CDAF model instead, the RBH becomes more luminous in the X-ray than in the 
optical ($f_{RASS} > f_{SDSS}$).  If we assume that $f_{SDSS}=5\times10^{-4}$, then
$N_{SDSS}$ within the EDR becomes $45f$.  Given $F_{RASS}^{min} = 10^{-14}$ erg cm$^{-2}$ s$^{-1}$
and $f_{RASS}=5\times10^{-2}$, $N_{RASS}$ within the EDR becomes $4500f$.  

The above estimates still do not represent the expected number of candidates until
a value of $f$ is specified.  The value of $\rho_{DM}$ represents the local value of the
galactic dark matter halo.  Recent microlensing work by \citet{Lasserre} indicates that the 
halo may be made of of compact objects if their mass is greater than 1 $M_\odot$.  \citet{Venkatesan}
place a limit of $f \lesssim 0.3 - 0.4$ by requiring that the RBH progenitor population 
of stars not overenrich the galaxy with metals.  

The next step in the program would be follow-up observations of the 57
remaining candidates.  Spectroscopic determination of stellar or QSO-like spectra would
rule out the candidate.  Higher resolution X-ray observations would confirm X-ray
emission.  Determination of variability either in optical or
X-ray, or any indication of proper motion would make the candidate source very
interesting.

Of the three methods of detecting isolated RBH mentioned earlier, only one
(microlensing) has possibly been successful.  This, however, does not decrease
the need for detecting accretion emission.  As microlensing uses the
magnification of light as an object passes between source and observer, it is
most sensitive to high-velocity black holes.  Since the accretion luminosity
scales as $L \propto \dot{M} \propto v^{-3}$, it would be expected that a RBH
detected in microlensing would not be detected through its emission [see
\citet{Revnivtsev}].  Thus these two methods are complementary, as those RBH
with low $v$ would not be detectable with microlensing but more likely through
accretion.

It should be noted that the three methods for detecting RBH's are not as
sensitive to primordial black holes.  Detection through supernova light curves
is obviously out.  Since $L \propto \dot{M} \propto M^2$, black holes with
masses much lighter than solar become undetectable through accretion emission.
Similarly for microlensing, the peak magnification scales as $A \propto
M^{1/2}$.

Finally we turn to limits on the contribution of RBHs to the local dark-matter
density. Of course any limit is only as reliable as the assumptions used to
derive it.  In this case, the major uncertainty is the source spectrum.
Nevertheless, if we assume that all 55 of the non-identified sources in Table
\ref{rocky1-47} are sources, then using equation (\ref{nobs}) with the
coefficient appropriate for the SDSS Early Data Release ($4.5 \times 10^4f$
rather than $10^6f$) the limit on the contribution of remnant black holes is
$10^{-3}$ of the dark matter density.  Of course there are many
assumptions that are behind this limit.  It assumes that the spectral energy
density is given by ADAF models, and that all possible SDSS sources would have
been seen in the RASS.  For these reasons, we view this work as a search for
isolated, stellar-mass black holes, rather than an attempt to place a limit on
their contribution to local dark matter.

During the writing of this paper, one of our X-ray selected candidates, \# 21, 
was observed 
spectroscopically to be a $z=0.927$ QSO.  We thank Mike Brotherton
and Paul Nandra for bringing this to our attention and providing the spectrum.

\section*{Acknowledgments}

We are grateful to Steve Kent for his assistance with sdssQT and other points
about SDSS data analysis; and to John Cannizzo, Brian Wilhite, John Beacom, Hugh Harris, 
Scott Anderson, Gordon Richards, and Wendy Freedman for useful 
discussions.  This work
was supported in part by the Department of Energy and by NASA (NAG5-10842) and
by NSF Grant PHY-0079251.  This research has made use of the SIMBAD database,
operated at CDS, Strasbourg, France, and of the NASA/IPAC Extragalactic
Database (NED) which is operated by the Jet Propulsion Laboratory, California
Institute of Technology, under contract with the National Aeronautics and Space
Administration.

Funding for the creation and distribution of the SDSS Archive has been provided
by the Alfred P. Sloan Foundation, the Participating Institutions, the National
Aeronautics and Space Administration, the National Science Foundation, the US
Department of Energy, the Japanese Monbukagakusho, and the Max Planck Society.
The SDSS Web site is http://www.sdss.org/.
The Participating Institutions are The University of Chicago, Fermilab, the 
Institute for Advanced Study, the Japan Participation Group, The Johns Hopkins 
University, Los Alamos National Laboratory, the Max-Planck-Institute for 
Astronomy (MPIA), the Max-Planck-Institute for Astrophysics (MPA), New Mexico 
State University, Princeton University, the United States Naval Observatory, 
and the University of Washington.


\clearpage

\clearpage

\clearpage

\clearpage

\clearpage

\clearpage

\clearpage

\clearpage

\clearpage

\begin{deluxetable}{cccccc}
\tabletypesize{\scriptsize}
\tablecaption{SDSS Filter Characteristics \label{sdsstable}}
\tablewidth{0pt}
\tablehead{
\colhead{Band} & 
\colhead{$\nu_\textrm{min}$ (10$^{14}$ Hz)} &
\colhead{$\bar{\nu}$ (10$^{14}$ Hz)  } &
\colhead{$\nu_\textrm{max}$ (10$^{14}$ Hz)} &
\colhead{95\% completeness limit} & 
\colhead{$F_{\alpha}^{min}$ (10$^{-15}$ ergs s$^{-1}$ cm$^{-2}$)} 
}
\startdata
$u$ & 8    & 8.57 & 9    & 22.0 & 0.89 \\
$g$ & 5.5  & 6.25 & 7    & 22.2 & 3.53 \\
$r$ & 4.25 & 4.80 & 5.25 & 22.2 & 2.66 \\
$i$ & 3.75 & 3.90 & 4.25 & 21.3 & 3.59 \\
$z$ & 3    & 3.30 & 3.5  & 20.5 & 1.39 \\
\enddata
\tablecomments{The average wavelengths, frequencies, completeness limits, and
corresponding limiting fluxes for the SDSS photometric filters.  The 95\% completeness limit is
for point sources, from \citet{Stoughton}.} 
\end{deluxetable}


\clearpage

\begin{deluxetable}{l|crcccccrcc}
\rotate
\tablecaption{X-ray selected remnant black hole candidates. \label{rocky1-47}}
\tablewidth{0pt}
\tablehead{\# &
\colhead{RA} & \colhead{dec}   & \colhead{$u^*$}  & \colhead{$g^*$} &
\colhead{$r^*$}  & \colhead{$i^*$} & \colhead{$z^*$} &
\colhead{RASS}     & \colhead{$\eta$}  }
\startdata
   1\tablenotemark{1} &  01  55 43.4 &  00  28 07.2 & 15.90 $\pm$  0.02 & 15.27 $\pm$  0.01 & 15.25 $\pm$  0.02 & 15.51 $\pm$  0.01 & 15.69 $\pm$  0.01 &   42 $\pm$   12 &1799.6 \\
   2 &  17  34 25.5 &  60  39 37.8 & 19.74 $\pm$  0.05 & 19.37 $\pm$  0.02 & 19.44 $\pm$  0.02 & 19.40 $\pm$  0.03 & 19.66 $\pm$  0.08 &    6 $\pm$    2\phn &1799.6 \\
   3\tablenotemark{2} &  13  11 06.5 &  00  35 10.1 & 18.37 $\pm$  0.01 & 18.04 $\pm$  0.02 & 17.78 $\pm$  0.01 & 17.57 $\pm$  0.02 & 17.32 $\pm$  0.02 &  105 $\pm$   24 & 128.5  \\
   4\tablenotemark{1} &  15  07 38.0 &  00  18 51.4 & 18.16 $\pm$  0.02 & 17.17
$\pm$  0.03 & 16.65 $\pm$  0.01 & 16.53 $\pm$  0.01 & 16.81 $\pm$  0.05 &   18
$\pm$    9\phn & 112.5  \\
   5\tablenotemark{2} &  00  10 47.5 &  00  19 00.4 & 18.92 $\pm$  0.02 & 18.75 $\pm$  0.01 & 18.77 $\pm$  0.02 & 18.74 $\pm$  0.02 & 18.73 $\pm$  0.06 &   18 $\pm$    9\phn & 105.9 \\
   6\tablenotemark{2} &  11  45 10.4 &  01  10 56.2 & 19.28 $\pm$  0.03 & 18.96 $\pm$  0.01 & 19.06 $\pm$  0.01 & 19.01 $\pm$  0.02 & 19.00 $\pm$  0.05 &   26 $\pm$   12 & 105.9 \\
   7\tablenotemark{2} &  11  52 46.6 &  00  24 40.0 & 17.63 $\pm$  0.01 & 17.53 $\pm$  0.02 & 17.56 $\pm$  0.01 & 17.53 $\pm$  0.01 & 17.47 $\pm$  0.02 &   27 $\pm$   11 & 105.9 \\
   8 &  17  09 21.6 &  57  06 24.0 & 21.85 $\pm$  0.18 & 21.00 $\pm$  0.04 & 20.11 $\pm$  0.02 & 19.61 $\pm$  0.03 & 19.13 $\pm$  0.06 &    9 $\pm$    3\phn & 105.9 \\
   9\tablenotemark{4} &  00  34 43.9 & $-$00  54 13.1 & 19.59 $\pm$  0.03 & 19.05 $\pm$  0.03 & 19.16 $\pm$  0.02 & 19.12 $\pm$  0.03 & 19.07 $\pm$  0.06 &   37 $\pm$   10 &  69.2 \\
  10\tablenotemark{2} &  17  19 36.7 &  60  47 48.1 & 18.84 $\pm$  0.03 & 18.73 $\pm$  0.01 & 18.74 $\pm$  0.01 & 18.80 $\pm$  0.02 & 18.72 $\pm$  0.04 &   14 $\pm$    4\phn &  45.0 \\
  11\tablenotemark{1} &  11  46 35.2 &  00  12 33.5 & 14.17 $\pm$  0.01 & 14.78 $\pm$  0.02 & 15.39 $\pm$  0.02 & 15.77 $\pm$  0.02 & 16.12 $\pm$  0.03 &   91 $\pm$   20 &  43.9  \\
  12\tablenotemark{2} &  16  49 31.1 &  64  21 31.0 & 19.14 $\pm$  0.03 & 18.89 $\pm$  0.02 & 18.92 $\pm$  0.02 & 18.88 $\pm$  0.02 & 18.90 $\pm$  0.05 &    4 $\pm$    2\phn &  35.3 \\
  13 &  15  26 14.5 & $-$00  44 45.2 & 18.27 $\pm$  0.02 & 18.17 $\pm$  0.01 & 18.15 $\pm$  0.02 & 18.23 $\pm$  0.01 & 18.19 $\pm$  0.02 &   17 $\pm$    7\phn &  33.3  \\
  14\tablenotemark{2} &  13  51 28.3 &  01  03 38.6 & 17.37 $\pm$  0.01 & 17.16 $\pm$  0.02 & 16.95 $\pm$  0.02 & 16.95 $\pm$  0.01 & 17.01 $\pm$  0.02 &   40 $\pm$   13 &  31.3  \\
  15 &  17  32 58.8 &  59  35 12.0 & 19.74 $\pm$  0.04 & 19.45 $\pm$  0.02 & 19.22 $\pm$  0.02 & 19.36 $\pm$  0.02 & 19.44 $\pm$  0.07 &    6 $\pm$    2\phn &  31.3 \\
  16\tablenotemark{2} &  17  00 35.4 &  63  25 22.7 & 18.15 $\pm$  0.02 & 18.03 $\pm$  0.02 & 18.00 $\pm$  0.06 & 18.05 $\pm$  0.02 & 18.09 $\pm$  0.03 &    4 $\pm$    2\phn &  31.3 \\
  17 &  10  18 27.1 & -00  00 08.5 & 19.54 $\pm$  0.03 & 19.35 $\pm$  0.01 & 19.19 $\pm$  0.02 & 19.25 $\pm$  0.02 & 19.42 $\pm$  0.06 &   26 $\pm$   10 &  31.3  \\
  18\tablenotemark{2} &  17  23 58.0 &  60  11 40.1 & 19.72 $\pm$  0.03 & 19.09 $\pm$  0.02 & 18.67 $\pm$  0.02 & 18.34 $\pm$  0.02 & 17.99 $\pm$  0.03 &   10 $\pm$    3\phn &  28.6 \\
  19\tablenotemark{2} &  11  32 45.6 &  00  34 27.8 & 18.28 $\pm$  0.02 & 17.82 $\pm$  0.01 & 17.40 $\pm$  0.01 & 17.09 $\pm$  0.02 & 16.77 $\pm$  0.01 &   19 $\pm$    9\phn &  24.7 \\
  20\tablenotemark{2} &  09  48 57.3 &  00  22 25.5 & 18.61 $\pm$  0.02 & 18.37 $\pm$  0.01 & 18.29 $\pm$  0.01 & 18.13 $\pm$  0.01 & 18.16 $\pm$  0.03 &   41 $\pm$   11 &  21.3 \\
  21 &  17  14 38.6 &  61  50 39.4 & 20.07 $\pm$  0.05 & 19.76 $\pm$  0.02 & 19.54 $\pm$  0.02 & 19.50 $\pm$  0.03 & 19.62 $\pm$  0.08 &    8 $\pm$    3\phn &  21.3 \\
  22\tablenotemark{2} &  17  10 30.2 &  60  23 47.6 & 18.01 $\pm$  0.02 & 17.76 $\pm$  0.02 & 17.54 $\pm$  0.01 & 17.30 $\pm$  0.02 & 17.38 $\pm$  0.02 &   12 $\pm$    4\phn &  21.3 \\
  23\tablenotemark{2} &  13  54 25.2 &  $-$00  13 58.0 & 16.82 $\pm$  0.02 & 16.65 $\pm$  0.01 & 16.59 $\pm$  0.01 & 16.44 $\pm$  0.01 & 16.45 $\pm$  0.02 &   39 $\pm$   13 &  21.3 \\
  24 &  11  10 34.5 & $-$01  05 17.5 & 20.44 $\pm$  0.07 & 20.15 $\pm$  0.03 & 19.96 $\pm$  0.03 & 19.73 $\pm$  0.03 & 19.80 $\pm$  0.10 &   16 $\pm$    9\phn &  21.3 \\
  25 &  00  35 34.2 & $-$00  25 48.8 & 19.51 $\pm$  0.03 & 19.35 $\pm$  0.02 & 19.14 $\pm$  0.02 & 19.11 $\pm$  0.02 & 19.35 $\pm$  0.08 &   20 $\pm$    8\phn &  21.3  \\
  26\tablenotemark{2} &  12  13 47.5 &  00  01 30.0 & 18.28 $\pm$  0.01 & 18.02 $\pm$  0.03 & 17.87 $\pm$  0.03 & 17.89 $\pm$  0.02 & 17.86 $\pm$  0.02 &   39 $\pm$   15 &  18.5  \\
  27 &  17  20 28.8 &  65  19 40.2 & 19.36 $\pm$  0.03 & 19.12 $\pm$  0.02 & 18.98 $\pm$  0.02 & 19.07 $\pm$  0.02 & 19.04 $\pm$  0.06 &    5 $\pm$    2\phn &  18.5  \\
  28 &  17  23 16.2 &  53  36 31.2 & 19.07 $\pm$  0.02 & 19.04 $\pm$  0.01 & 18.98 $\pm$  0.02 & 19.05 $\pm$  0.04 & 19.18 $\pm$  0.07 &    8 $\pm$    3\phn &  16.4 \\
  29\tablenotemark{2} &  17  08 32.2 &  62  42 05.8 & 19.12 $\pm$  0.02 & 18.77
$\pm$  0.01 & 18.55 $\pm$  0.03 & 18.57 $\pm$  0.02 & 18.66 $\pm$  0.04 &    9
$\pm$    3\phn &  14.5 \\
  30\tablenotemark{2} &  12  10 16.1 &  00  12 04.9 & 17.03 $\pm$  0.01 & 16.96 $\pm$  0.02 & 16.92 $\pm$  0.01 & 16.86 $\pm$  0.01 & 16.92 $\pm$  0.04 &   18. $\pm$   9\phn  &  12.1 \\
  31\tablenotemark{2} &  03  09 11.6 &  00  23 58.9 & 16.94 $\pm$  0.02 & 16.70 $\pm$  0.01 & 16.84 $\pm$  0.01 & 16.89 $\pm$  0.01 & 17.02 $\pm$  0.02 &   59 $\pm$   18 &   9.0  \\
  32\tablenotemark{2} &  11  37 49.8 &  00  27 35.3 & 17.50 $\pm$  0.01 & 17.26 $\pm$  0.02 & 17.24 $\pm$  0.03 & 17.20 $\pm$  0.01 & 17.08 $\pm$  0.03 &   33 $\pm$   13 &   8.4  \\
  33\tablenotemark{2} &  17  15 08.1 &  55  29 25.0 & 17.93 $\pm$  0.02 & 17.76 $\pm$  0.02 & 17.63 $\pm$  0.02 & 17.62 $\pm$  0.02 & 17.59 $\pm$  0.02 &   11 $\pm$    4\phn &   8.4  \\
  34 &  17  20 52.3 &  57  55 13.2 & 20.52 $\pm$  0.06 & 20.21 $\pm$  0.05 & 19.98 $\pm$  0.05 & 19.93 $\pm$  0.09 & 19.81 $\pm$  0.12 &   14 $\pm$    4\phn &   8.4 \\
  35 &  17  22 40.1 &  61  05 60.0 & 19.37 $\pm$  0.03 & 19.10 $\pm$  0.01 & 19.04 $\pm$  0.02 & 19.01 $\pm$  0.02 & 18.86 $\pm$  0.05 &   12 $\pm$    3\phn &   8.4  \\
  36\tablenotemark{4} &  00  59 18.2 &  00  25 19.7 & 18.45 $\pm$  0.02 & 18.06 $\pm$  0.02 &
18.03 $\pm$  0.01 & 17.96 $\pm$  0.01 & 17.90 $\pm$  0.03 &   46 $\pm$   14 &
7.7 \\
  37 &  17  45 04.4 &  53  20 27.5 & 19.35 $\pm$  0.03 & 18.86 $\pm$  0.02 & 18.73 $\pm$  0.01 & 18.62 $\pm$  0.02 & 18.43 $\pm$  0.03 &   11 $\pm$    3\phn &   7.7  \\
  38 &  01  46 01.7 & $-$00  21 22.0 & 20.41 $\pm$  0.08 & 20.01 $\pm$  0.02 & 19.84 $\pm$  0.02 & 19.68 $\pm$  0.02 & 19.46 $\pm$  0.07 &   18 $\pm$    8\phn &   7.7 \\
  39 &  11  54 12.0 &  01  00 57.6 & 21.15 $\pm$  0.16 & 20.45 $\pm$  0.04 & 20.14 $\pm$  0.03 & 19.95 $\pm$  0.07 & 19.68 $\pm$  0.12 &   33 $\pm$   12 &   5.7  \\
  40\tablenotemark{2} &  17  14 30.1 &  61  57 46.6 & 19.90 $\pm$  0.05 & 19.39 $\pm$  0.02 & 19.09 $\pm$  0.02 & 19.00 $\pm$  0.02 & 18.86 $\pm$  0.05 &   17 $\pm$    4\phn &   5.2 \\
  41 &  17  33 49.7 &  58  43 57.6 & 21.74 $\pm$  0.18 & 21.17 $\pm$  0.04 & 20.70 $\pm$  0.04 & 20.57 $\pm$  0.05 & 20.42 $\pm$  0.15 &    7 $\pm$    2\phn &   5.2 \\
  42\tablenotemark{2} &  12  03 46.6 & $-$00  17 23.1 & 19.98 $\pm$  0.05 & 19.58 $\pm$  0.03 & 19.14 $\pm$  0.02 & 19.08 $\pm$  0.02 & 19.06 $\pm$  0.07 &   26 $\pm$   11 &   5.2  \\
  43 &  11  04 54.8 & $-$01  08 53.4 & 20.86 $\pm$  0.10 & 20.38 $\pm$  0.03 & 19.93 $\pm$  0.03 & 19.76 $\pm$  0.04 & 19.61 $\pm$  0.08 &   16 $\pm$    8\phn &   5.2  \\
  44\tablenotemark{2} &  17  09 56.0 &  57  32 25.5 & 18.63 $\pm$  0.02 & 18.26 $\pm$  0.01 & 18.21 $\pm$  0.01 & 18.09 $\pm$  0.02 & 18.10 $\pm$  0.04 &   28 $\pm$    5\phn &   5.0 \\
  45 &  13  31 11.1 &  01  00 12.3 & 20.61 $\pm$  0.06 & 19.72 $\pm$  0.03 & 19.51 $\pm$  0.03 & 19.43 $\pm$  0.04 & 19.32 $\pm$  0.07 &   30 $\pm$   12 &   1.1  \\
  46 &  12  14 42.0 &  00  40 17.5 & 20.63 $\pm$  0.05 & 19.72 $\pm$  0.02 & 19.49 $\pm$  0.02 & 19.39 $\pm$  0.02 & 19.37 $\pm$  0.05 &   24 $\pm$   11 &   1.1  \\
  47 &  17  25 32.1 &  57  16 35.4 & 20.83 $\pm$  0.09 & 19.91 $\pm$  0.02 & 19.70 $\pm$  0.02 & 19.51 $\pm$  0.03 & 19.43 $\pm$  0.07 &   27 $\pm$    6\phn &   1.1\\
  48\tablenotemark{3} &  17  31 00.4 &  57  22 12.4 & 19.10 $\pm$  0.03 & 18.16 $\pm$  0.02 & 17.94 $\pm$  0.01 & 17.81 $\pm$  0.02 & 17.62 $\pm$  0.02 &    8 $\pm$    3\phn &   1.1 \\
  49 &  15  24 37.1 &  00  18 46.2 & 18.94 $\pm$  0.03 & 18.04 $\pm$  0.02 & 17.82 $\pm$  0.02 & 17.77 $\pm$  0.01 & 17.71 $\pm$  0.03 &   23 $\pm$   11 &   1.1  \\
  50\tablenotemark{1} &  15  13 45.0 &  00  18 20.7 & 19.57 $\pm$  0.03 & 18.72
$\pm$  0.01 & 18.51 $\pm$  0.01 & 18.43 $\pm$  0.01 & 18.36 $\pm$  0.03 &   23
$\pm$   10 &   1.1 \\
  51 &  12  50 28.1 & $-$00  46 56.5 & 18.64 $\pm$  0.02 & 17.74 $\pm$  0.01 & 17.50 $\pm$  0.03 & 17.39 $\pm$  0.02 & 17.36 $\pm$  0.02 &   45 $\pm$   19 &   1.1 \\
  52 &  12  14 41.4 &  00  40 32.9 & 21.41 $\pm$  0.09 & 20.66 $\pm$  0.02 & 20.41 $\pm$  0.03 & 20.28 $\pm$  0.03 & 20.39 $\pm$  0.11 &   24 $\pm$   11 &   0.9  \\
  53\tablenotemark{2} &  12  15 25.1 &  00  53 16.9 & 20.94 $\pm$  0.10 & 20.34 $\pm$  0.03 & 19.94 $\pm$  0.03 & 19.93 $\pm$  0.04 & 19.95 $\pm$  0.12 &   26 $\pm$   11 &   0.9 \\
  54 &  16  59 50.8 &  62  38 45.5 & 21.03 $\pm$  0.11 & 20.20 $\pm$  0.02 & 19.88 $\pm$  0.02 & 19.64 $\pm$  0.03 & 19.71 $\pm$  0.10 &    6 $\pm$    2\phn &   0.9 \\
  55 &  23  47 25.2 & $-$01  06 36.0 & 20.35 $\pm$  0.08 & 19.51 $\pm$  0.02 & 19.11 $\pm$  0.02 & 18.97 $\pm$  0.02 & 18.99 $\pm$  0.06 &   48 $\pm$   14 &   0.9    \\
  56\tablenotemark{1} &  10  47 20.6 & $-$00  41 48.3 & 21.32 $\pm$  0.15 & 20.50 $\pm$  0.04 & 20.17 $\pm$  0.04 & 20.03 $\pm$  0.04 & 20.25 $\pm$  0.16 &   42 $\pm$   12 &   0.9  \\
  57 &  17  11 22.3 &  58  04 60.0 & 20.94 $\pm$  0.08 & 20.15 $\pm$  0.03 & 19.86 $\pm$  0.03 & 19.75 $\pm$  0.02 & 19.58 $\pm$  0.08 &   10 $\pm$    4\phn &   0.8  \\
  58 &  17  15 24.3 &  55  00 14.1 & 21.78 $\pm$  0.19 & 21.03 $\pm$  0.05 & 20.63 $\pm$  0.04 & 20.43 $\pm$  0.05 & 20.23 $\pm$  0.14 &   10 $\pm$    4\phn &   0.8 \\
  59 &  17  15 34.2 &  63  23 45.5 & 20.45 $\pm$  0.08 & 19.64 $\pm$  0.02 & 19.38 $\pm$  0.02 & 19.27 $\pm$  0.02 & 19.18 $\pm$  0.07 &    6 $\pm$    2\phn &   0.8 \\
  60 &  17  06 05.9 &  64  38 20.2 & 20.76 $\pm$  0.09 & 19.98 $\pm$  0.03 & 19.70 $\pm$  0.02 & 19.58 $\pm$  0.03 & 19.56 $\pm$  0.09 &    6 $\pm$    2\phn &   0.8 \\
  61 &  17  31 34.2 &  59  13 52.7 & 18.49 $\pm$  0.02 & 17.65 $\pm$  0.01 & 17.37 $\pm$  0.01 & 17.30 $\pm$  0.01 & 17.29 $\pm$  0.02 &    4 $\pm$    2\phn &   0.8 \\
  62 &  15  16 57.1 & $-$00  37 24.6 & 19.09 $\pm$  0.02 & 18.24 $\pm$  0.02 & 17.95 $\pm$  0.01 & 17.84 $\pm$  0.02 & 17.81 $\pm$  0.02 &  176 $\pm$   32 &   0.8  \\
  63 &  14  37 37.5 & $-$00  20 07.5 & 22.17 $\pm$  0.19 & 21.36 $\pm$  0.04 & 20.88 $\pm$  0.04 & 20.77 $\pm$  0.04 & 20.68 $\pm$  0.16 &   85 $\pm$   23 &   0.8  \\
  64 &  02  25 07.9 & $-$00  35 33.0 & 19.44 $\pm$  0.03 & 18.83 $\pm$  0.01 & 18.48 $\pm$  0.02 & 18.33 $\pm$  0.02 & 18.12 $\pm$  0.04 &   70 $\pm$   20 &   0.8   \\
  65 &  01  04 14.9 & $-$00  24 34.0 & 21.41 $\pm$  0.16 & 20.57 $\pm$  0.03 & 20.14 $\pm$  0.03 & 19.94 $\pm$  0.03 & 19.77 $\pm$  0.12 &   21 $\pm$    9\phn &   0.8 \\
  66 &  17  24 11.7 &  57  17 28.4 & 20.91 $\pm$  0.09 & 20.01 $\pm$  0.03 & 19.62 $\pm$  0.02 & 19.46 $\pm$  0.02 & 19.48 $\pm$  0.07 &    8 $\pm$    3\phn &   0.6    \\
  67 &  16  49 36.7 &  64  28 15.2 & 21.43 $\pm$  0.13 & 20.43 $\pm$  0.03 & 19.96 $\pm$  0.02 & 19.72 $\pm$  0.03 & 19.86 $\pm$  0.11 &   41 $\pm$    6\phn &   0.6  \\
  68 &  16  58 53.7 &  63  27 51.8 & 20.94 $\pm$  0.09 & 20.08 $\pm$  0.02 & 19.77 $\pm$  0.04 & 19.77 $\pm$  0.03 & 19.85 $\pm$  0.11 &    3 $\pm$    1\phn &   0.6  \\
  69 &  17  24 08.2 &  64  49 24.4 & 17.62 $\pm$  0.02 & 16.61 $\pm$  0.01 & 16.24 $\pm$  0.02 & 16.10 $\pm$  0.01 & 16.10 $\pm$  0.02 &    5 $\pm$    2\phn &   0.6  \\
  70 &  14  37 30.6 & $-$00  21 16.5 & 18.28 $\pm$  0.01 & 17.35 $\pm$  0.01 & 17.00 $\pm$  0.01 & 16.88 $\pm$  0.01 & 16.89 $\pm$  0.02 &   26 $\pm$   11 &   0.6  \\
  71 &  11  04 54.7 & $-$01  09 11.9 & 20.00 $\pm$  0.06 & 19.09 $\pm$  0.02 & 18.78 $\pm$  0.02 & 18.73 $\pm$  0.03 & 18.76 $\pm$  0.04 &   16 $\pm$    8\phn &   0.6  \\
  72 &  00  17 25.5 & $-$01  11 51.4 & 19.56 $\pm$  0.04 & 18.56 $\pm$  0.01 & 18.19 $\pm$  0.02 & 18.10 $\pm$  0.01 & 18.11 $\pm$  0.03 &   22 $\pm$   10 &   0.6   \\
  73 &  01  31 44.7 &  00  33 04.9 & 19.80 $\pm$  0.04 & 18.92 $\pm$  0.02 & 18.58 $\pm$  0.01 & 18.47 $\pm$  0.02 & 18.40 $\pm$  0.03 &   46 $\pm$   14 &   0.4   \\
  74 &  13  56 15.4 &  00  03 58.1 & 18.93 $\pm$  0.02 & 17.95 $\pm$  0.01 & 17.64 $\pm$  0.02 & 17.51 $\pm$  0.01 & 17.50 $\pm$  0.02 &   24 $\pm$   11 &   0.4  \\
  75 &  17  23 57.3 &  58  33 08.1 & 20.78 $\pm$  0.10 & 19.81 $\pm$  0.02 & 19.44 $\pm$  0.02 & 19.33 $\pm$  0.02 & 19.25 $\pm$  0.06 &   13 $\pm$    4\phn &   0.4  \\
  76 &  17  27 00.6 &  58  19 17.1 & 20.78 $\pm$  0.10 & 19.86 $\pm$  0.03 & 19.48 $\pm$  0.02 & 19.24 $\pm$  0.02 & 19.17 $\pm$  0.05 &    8 $\pm$    4\phn &   0.4   \\
  77 &  17  24 03.1 &  52  53 45.6 & 21.53 $\pm$  0.12 & 20.51 $\pm$  0.03 & 20.12 $\pm$  0.03 & 19.99 $\pm$  0.04 & 19.86 $\pm$  0.11 &   10 $\pm$    4\phn &   0.4  \\
  78 &  15  35 58.5 &  00  03 39.6 & 17.44 $\pm$  0.01 & 16.56 $\pm$  0.01 & 16.28 $\pm$  0.01 & 16.20 $\pm$  0.01 & 16.18 $\pm$  0.02 &   22 $\pm$    8\phn &   0.4    \\
  79 &  15  32 53.3 & $-$00  46 02.5 & 19.39 $\pm$  0.04 & 18.38 $\pm$  0.02 & 17.96 $\pm$  0.02 & 17.75 $\pm$  0.01 & 17.69 $\pm$  0.02 &   27 $\pm$   11 &   0.4   \\
  80 &  14  31 19.3 & $-$00  54 37.2 & 19.61 $\pm$  0.03 & 18.62 $\pm$  0.01 & 18.33 $\pm$  0.01 & 18.25 $\pm$  0.01 & 18.24 $\pm$  0.03 &   40 $\pm$   14 &   0.4  \\
  81 &  13  04 27.0 & $-$00  35 41.6 & 19.22 $\pm$  0.02 & 18.34 $\pm$  0.02 & 17.99 $\pm$  0.02 & 17.81 $\pm$  0.01 & 17.73 $\pm$  0.02 &   30 $\pm$   14 &   0.4   \\
  82 &  12  50 23.6 & $-$00  47 49.0 & 19.42 $\pm$  0.03 & 18.54 $\pm$  0.02 & 18.17 $\pm$  0.03 & 17.96 $\pm$  0.02 & 17.92 $\pm$  0.03 &   45 $\pm$   19 &   0.4  \\
  83 &  13  14 41.2 & $-$01  07 01.5 & 18.73 $\pm$  0.02 & 17.78 $\pm$  0.02 & 17.44 $\pm$  0.02 & 17.32 $\pm$  0.01 & 17.26 $\pm$  0.02 &   32 $\pm$   14 &   0.4 \\
  84 &  11  40 24.7 & $-$00  59 26.7 & 19.79 $\pm$  0.03 & 18.88 $\pm$  0.01 & 18.60 $\pm$  0.01 & 18.48 $\pm$  0.01 & 18.43 $\pm$  0.03 &   60 $\pm$   22 &   0.4  \\
  85 &  11  40 28.4 & $-$00  15 51.2 & 19.26 $\pm$  0.03 & 18.32 $\pm$  0.02 & 17.98 $\pm$  0.01 & 17.85 $\pm$  0.02 & 17.79 $\pm$  0.03 &   36 $\pm$   15 &   0.4   \\
  86\tablenotemark{1} &  12  12 22.8 &  00  25 46.9 & 19.90 $\pm$  0.04 & 18.97 $\pm$  0.02 & 18.77 $\pm$  0.01 & 18.69 $\pm$  0.03 & 18.69 $\pm$  0.05 &   20 $\pm$    9\phn &   0.4 \\
  87 &  16  56 22.4 &  64  35 43.6 & 20.73 $\pm$  0.08 & 19.72 $\pm$  0.03 & 19.49 $\pm$  0.02 & 19.41 $\pm$  0.03 & 19.44 $\pm$  0.08 &    4 $\pm$    2\phn &   0.4\\
\enddata
\tablenotetext{1}{Object identified as a star in the SDSS spectroscopic survey.}
\tablenotetext{2}{Object identified as a low redshift ($z \lesssim 2.3$) QSO in the SDSS spectroscopic survey EDR database.}
\tablenotetext{3}{Object identified as a high redshift ($z \gtrsim 2.3$) QSO in the SDSS spectroscopic survey EDR database.}
\tablenotetext{4}{Object identified as a QSO in \citet{Richards}.}
\tablecomments{Positions are J2000.
SDSS magnitudes $u^*$, $g^*$, $r^*$, $i^*$, $z^*$ are reddening corrected
point-spread-function (PSF) magnitudes. The RASS count rate is in counts ksec$^{-1}$.
All errors are 1-$\sigma$.  The overdensity $\eta$ is defined in 
equation (\ref{overdensity}), and is used to rank those candidates most likely
to be RBHs.
}
\end{deluxetable}

\begin{deluxetable}{l|crcccccccc}
\rotate
\tablecaption{Spectroscopically selected remnant black hole candidates. \label{spectrocandidates}}
\tablewidth{0pt}
\tablehead{\# &
\colhead{RA} & \colhead{dec}   & \colhead{$u^*$}  & \colhead{$g^*$} &
\colhead{$r^*$}  & \colhead{$i^*$} & \colhead{$z^*$} &
\colhead{Plate}     & \colhead{MJD} & \colhead{Fiber}}
\startdata
1\tablenotemark{1}&  01  00 58.2 & -00 55 47.9 & 19.68 $\pm$  0.03 & 19.17 $\pm$  0.02 & 18.76 $\pm$  0.02 & 18.36 $\pm$  0.02 & 18.10 $\pm$  0.03 & 395 &   51783 &    18  \\
2 &  01  27 23.6 & -00 46 30.1 & 18.79 $\pm$  0.02 & 18.59 $\pm$  0.01 & 18.71 $\pm$  0.01 & 18.76 $\pm$  0.02 & 18.85 $\pm$  0.05 & 399 &   51817 &    99  \\
3 &  03  16 42.7 & -00  08 16.7 & 18.83 $\pm$  0.02 & 18.85 $\pm$  0.02 & 18.98 $\pm$  0.02 & 19.20 $\pm$  0.02 & 19.45 $\pm$  0.10 & 412 &   51931 &    21  \\
4 &  03  44 01.4 & -00 12 21.0 & 19.26 $\pm$  0.04 & 19.27 $\pm$  0.01 & 19.49 $\pm$  0.02 & 19.71 $\pm$  0.03 & 19.89 $\pm$  0.14 & 416 &   51811 &   111  \\
5 &  03  42 26.3 & -00 14 09.9 & 20.28 $\pm$  0.08 & 20.13 $\pm$  0.03 & 19.92 $\pm$  0.04 & 19.93 $\pm$  0.04 & 20.07 $\pm$  0.15 & 416 &   51811 &   154  \\
6 &  03  33 57.2 & -00 11 06.1 & 19.78 $\pm$  0.05 & 19.64 $\pm$  0.02 & 19.55 $\pm$  0.02 & 19.47 $\pm$  0.02 & 19.42 $\pm$  0.10 & 415 &   51810 &   174  \\
7 &  10  32 43.3 & -00 32 43.6 & 19.56 $\pm$  0.03 & 19.86 $\pm$  0.02 & 20.18 $\pm$  0.02 & 20.52 $\pm$  0.04 & 20.71 $\pm$  0.18 & 273 &   51957 &   163  \\
8\tablenotemark{1} &  12  25 19.9 & -01  07 36.9 & 20.11 $\pm$  0.04 & 19.47 $\pm$  0.02 & 19.10 $\pm$  0.01 & 19.11 $\pm$  0.02 & 19.11 $\pm$  0.06 & 289 &   51990 &   250  \\
9 &  15  24 40.1 &  00  32 52.1 & 19.47 $\pm$  0.03 & 19.02 $\pm$  0.01 & 18.73 $\pm$  0.02 & 18.58 $\pm$  0.01 & 18.43 $\pm$  0.03 & 313 &   51673 &   463  \\
10 &  17  43 52.5 &  54  54 38.8 & 20.50 $\pm$  0.08 & 20.22 $\pm$  0.03 & 20.37 $\pm$  0.03 & 20.37 $\pm$  0.05 & 20.39 $\pm$  0.16 & 360 &   51816 &   633  \\
11 &  17  11 01.5 &  65  45 49.9 & 19.18 $\pm$  0.03 & 18.79 $\pm$  0.02 & 18.68 $\pm$  0.02 & 18.73 $\pm$  0.02 & 18.90 $\pm$  0.05 & 350 &   51691 &   367  \\
12 &  17  33 27.3 &  58  54 39.8 & 19.95 $\pm$  0.04 & 19.88 $\pm$  0.02 & 19.98 $\pm$  0.02 & 20.10 $\pm$  0.04 & 20.16 $\pm$  0.12 & 366 &   52017 &   582  \\
13 &  17  09 27.5 &  62  29 01.5 & 19.14 $\pm$  0.03 & 18.91 $\pm$  0.01 & 18.84 $\pm$  0.03 & 18.97 $\pm$  0.04 & 19.02 $\pm$  0.05 & 351 &   51780 &   580  \\
14 &  17  24 11.5 &  58  37 10.9 & 20.70 $\pm$  0.09 & 20.36 $\pm$  0.02 & 20.34 $\pm$  0.04 & 20.30 $\pm$  0.04 & 20.23 $\pm$  0.13 & 366 &   52017 &   280  \\
15 &  17  22 28.9 &  58  40 10.9 & 19.76 $\pm$  0.04 & 19.79 $\pm$  0.02 & 20.11 $\pm$  0.03 & 20.32 $\pm$  0.05 & 20.40 $\pm$  0.16 & 366 &   52017 &   434  \\
16 &  17  33 42.9 &  55  44 19.1 & 18.86 $\pm$  0.02 & 18.56 $\pm$  0.02 & 18.57 $\pm$  0.01 & 18.64 $\pm$  0.01 & 18.71 $\pm$  0.04 & 360 &   51816 &   380  \\
17 &  17  24 00.7 &  57  35 38.2 & 20.22 $\pm$  0.05 & 20.21 $\pm$  0.02 & 20.40 $\pm$  0.04 & 20.47 $\pm$  0.05 & 20.68 $\pm$  0.17 & 366 &   52017 &   248  \\
18\tablenotemark{4}&  13  11 06.5 &  00  35 10.1 & 18.37 $\pm$  0.01 & 18.04 $\pm$  0.02 & 17.78 $\pm$  0.01 & 17.57 $\pm$  0.02 & 17.32 $\pm$  0.02 & 294 &   51986 &   629  \\
19 &  12  55 59.6 &  00  51 06.0 & 20.54 $\pm$  0.06 & 20.15 $\pm$  0.02 & 20.05 $\pm$  0.02 & 20.02 $\pm$  0.03 & 19.90 $\pm$  0.09 & 293 &   51689 &   372  \\
20 &  14  44 54.6 &  00  42 24.5 & 21.34 $\pm$  0.10 & 20.67 $\pm$  0.03 & 20.22 $\pm$  0.02 & 20.05 $\pm$  0.04 & 19.82 $\pm$  0.08 & 308 &   51662 &   414  \\
21 &  11  47 37.9 &  00  13 01.0 & 21.91 $\pm$  0.15 & 21.11 $\pm$  0.03 & 20.35 $\pm$  0.02 & 19.84 $\pm$  0.03 & 19.25 $\pm$  0.06 & 283 &   51959 &   543  \\
22 &  11  10 07.6 &  01  10 41.6 & 18.03 $\pm$  0.02 & 17.51 $\pm$  0.02 & 17.36 $\pm$  0.02 & 17.32 $\pm$  0.01 & 17.35 $\pm$  0.02 & 278 &   51900 &   523  \\
23\tablenotemark{2}  &  03  38 10.9 &  00  56 17.7 & 20.19 $\pm$  0.08 & 19.25 $\pm$  0.02 & 18.40 $\pm$  0.01 & 18.14 $\pm$  0.01 & 18.28 $\pm$  0.03 & 415 &   51810 &   617  \\
24\tablenotemark{3}&  03  33 20.4 &  00  07 20.6 & 16.56 $\pm$  0.02 & 16.19 $\pm$  0.01 & 16.17 $\pm$  0.01 & 16.18 $\pm$  0.01 & 16.37 $\pm$  0.01 & 415 &   51810 &   492  \\
25 &  01  07 48.2 &  01  02 40.7 & 19.09 $\pm$  0.03 & 18.70 $\pm$  0.02 & 18.49 $\pm$  0.02 & 18.50 $\pm$  0.01 & 18.50 $\pm$  0.04 & 396 &   51816 &   571  \\
26 &  02  51 11.7 &  00  29 15.1 & 19.38 $\pm$  0.04 & 19.17 $\pm$  0.02 & 19.31 $\pm$  0.02 & 19.52 $\pm$  0.03 & 19.60 $\pm$  0.09 & 410 &   51816 &   347  \\
27\tablenotemark{2} &  02  58 29.0 &  00  15 26.1 & 20.39 $\pm$  0.07 & 20.15 $\pm$  0.02 & 19.97 $\pm$  0.03 & 19.87 $\pm$  0.03 & 19.65 $\pm$  0.14 & 410 &   51816 &   559  \\
28 &  01  47 33.6 &  00  03 23.3 & 18.68 $\pm$  0.02 & 18.08 $\pm$  0.01 & 17.70 $\pm$  0.01 & 17.40 $\pm$  0.02 & 17.23 $\pm$  0.02 & 402 &   51793 &   400  \\
29 &  02  13 03.8 &  00  38 11.9 & 21.04 $\pm$  0.11 & 20.64 $\pm$  0.03 & 20.13 $\pm$  0.03 & 19.75 $\pm$  0.03 & 19.46 $\pm$  0.07 & 405 &   51816 &   469  \\
\enddata
\tablenotetext{1}{Objects identified as QSOs in \citet{Veron}.}
\tablenotetext{2}{Objects identified as QSOs in \citet{Schneider}.}
\tablenotetext{3}{Object identified as a DB WD in \citet{Reimers}.}
\tablenotetext{4}{This object is also X-ray selected object \#3.}
\tablecomments{Positions are J2000.
SDSS magnitudes $u^*$, $g^*$, $r^*$, $i^*$, $z^*$ are reddening corrected
point-spread-function (PSF) magnitudes.  All errors are 1-$\sigma$.  MJD is the Modified Julian Date of
when the spectra was taken.
}
\end{deluxetable}

\clearpage
\begin{deluxetable}{l|rrcccc}
\tablecaption{Proper Motions \label{propermotions}}
\tablewidth{0pt}
\tablehead{\# &
\colhead{$\pi$} & \colhead{$\delta$}   & \colhead{blue}  & \colhead{red} & \colhead{$g^*$} & \colhead{$r^*$} }
\startdata
X-1\tablenotemark{1}  &  62.6 &  2.82 & 18.2 & 16.5 & 15.38 & 15.32 \\
X-39 &  24.3 &  1.15 & 19.6 & 18.9 & 20.56 & 20.25 \\
X-44\tablenotemark{2} &  31.6 &  1.42 & 17.9 & 18.3 & 18.40 & 18.33 \\
X-70 &  32.6 &  1.43 & 17.0 & 16.7 & 17.49 & 17.10 \\
X-72 &  25.2 &  1.19 & 18.4 & 17.8 & 18.77 & 18.34 \\
X-81 &  34.9 &  1.50 & 18.3 & 17.9 & 18.37 & 18.02 \\
X-85 &  26.8 &  1.26 & 18.1 & 17.7 & 18.47 & 18.11 \\
S-2  &  97.9 &  4.30 & 17.8 & 18.6 & 18.67 & 18.76 \\
S-11 &  80.4 &  3.77 & 18.6 & 18.8 & 18.89 & 18.76 \\
S-22 & 406.6 & 17.86 & 16.9 & 17.1 & 17.63 & 17.43 \\
S-24\tablenotemark{3} &  93.5 &  4.18 & 16.2 & 16.2 & 16.52 & 16.41 \\
S-25 & 177.7 &  7.80 & 18.0 & 18.3 & 18.83 & 18.59 \\
\enddata
\tablenotetext{1}{Object identified as a CV.}
\tablenotetext{2}{Object identified as a QSO.}
\tablenotetext{3}{Object identified as a WD.}
\tablecomments{X refers to X-ray selected, S refers to spectroscopically
selected.  Proper motion $\pi$ is measured in mas yr$^{-1}$, and computed from
the separation $\delta$ in arcseconds and the elapsed time between the POSS and SDSS.
SDSS magnitudes are PSF and are not corrected for reddening.}
\end{deluxetable}


\begin{thebibliography}{99}

\bibitem[Abramowicz et al.(1995)]{AbramowiczChen}
Abramowicz, M. A., Chen, X., Kato, S., Lasota, J.-P., \& Regev, O., 1995,
\apj, 438, L37

\bibitem[Abramowicz \& Igumenshchev(2001)]{Abramowicz}
Abramowicz, M. A., \& Igumenshchev, I. V., 2001, \apj, 554, L53

\bibitem[Agol \& Kamionkowski(2002)]{Agol}
Agol, E. \& Kamionkowski, M., 2002, \mnras, 334, 553

\bibitem[Armitage \& Natarajan(1999)]{Armitage}
Armitage, P. J., \& Natarajan, P., 1999, \apj, 523, L7

\bibitem[Balberg \& Shapiro(2001)]{Balberg}
Balberg, S., \& Shapiro, S. L., 2001, \apj, 556, 944

\bibitem[Ball, Narayan \& Quataert(2001)]{Ball}
Ball, G. H., Narayan, R., \& Quataert, E., 2001, \apj, 552, 221

\bibitem[Beacom, Boyd \& Mezzacappa(2001)]{Beacom}
Beacom, J.F, Boyd, R.N., and Mezzacappa, A., 2001, \prd, 63, 073011

\bibitem[Begelman \& Meier(1982)]{Begelman}
Begelman, M.C. \& Meier, D.L., 1982 \apj, 253, 873

\bibitem[Bennett et al.(2002)]{Bennett}
Bennett, D.P, et al., 2002, \apj, 579, 639

\bibitem[Blandford \& Znajek(1977)]{Blandford}
Blandford, R. D., \& Znajek, R. L., 1977, \mnras, 179, 433

\bibitem[Bondi(1952)]{Bondi}
Bondi, H., 1952, \mnras, 112, 195

\bibitem[Campana \& Pardi(1993)]{Campana}
Campana, S. \& Pardi, M.C., 1993 \aap, 277, 477

\bibitem[Chakrabarti(1996)]{Chakrabarti}
Chakrabarti, S. K., 1996, \physrep, 266, 229

\bibitem[Ebisuzaki et al.(2001)]{Ebisuzaki}
Ebisuzaki, T., et al., 2001, \apjl, 562, L19 

\bibitem[Fan(1999)]{Fan}
Fan, X., 1999, \aj, 117, 2528

\bibitem[Fryer \& Kalogera(2001)]{Fryer}
Fryer, C.L., \& Kalogera, V., 2001, \apj, 554, 548

\bibitem[Fryxell \& Taam(1988)]{Fryxell}
Fryxell, B. A., \& Taam, R. E., 1988, \apj, 335, 862

\bibitem[Fujita at al.(1998)]{Fujita}
Fujita, Y., Inoue, S., Nakamura, T., Manmoto, T., \& Nakamura, K. E., 1998,
\apj, 95, L85


\bibitem[Gates, Gyuk \& Turner(1995)]{Gates}
Gates, E. I., Gyuk, G., \& Turner, M. S., \apjl, 449, L123

\bibitem[Grindlay(1978)]{Grindlay}
Grindlay, J. E., 1978, \apj, 221, 234


\bibitem[Hansen \& Phinney(1997)]{Hansen}
Hansen, B. M. S. \& Phinney, E. S., 1997, \mnras, 291, 569.

\bibitem[Heckler \& Kolb(1996)]{Heckler}
Heckler, A. F., \& Kolb, E. W., 1996, \apjl, 472, L85 


\bibitem[Ichimaru(1977)]{Ichimaru} 
Ichimaru, S. 1977, \apj, 214, 840

\bibitem[Igumenshchev \& Narayan(2002)]{Igumenshchev}
Igumenshchev, I. V., \& Narayan, R., 2002, \apj, 566, 137

\bibitem[Ipser \& Price(1982)]{Ipser}
Ipser, J. R., \& Price, R. H., 1982, \apj, 255, 654


\bibitem[Kaaret et al.(2001)]{Kaaret}
Kaaret, P., Prestwich, A.H., Zezas, A., Murray S.S., Kim D.-W., Kilgard, R.E.,
Schlegel, E.M., and Ward, M.J., 2001, \mnras, 321, L29

\bibitem[Kormendy \& Richstone(1995)]{Kormendy}
Kormendy, J., and Richstone, D. 1995, \araa, 33, 581


\bibitem[Lasserre et al.(2000)]{Lasserre}
Laserre, T., et al., 2000, \aap, 355, L39

\bibitem[Manmoto, Mineshige \& Kusunose(1997)]{Manmoto}
Manmoto, T., Mineshige, S., \& Kusunose, M., 1997, \apj, 489, 791

\bibitem[Mao et al.(2002)]{Mao}
Mao, S., et al, 2002, \mnras, 329, 349

\bibitem[Matsumoto \& Tsuru(1999)]{Matsumoto99} 
Matsumoto, H., and Tsuru, T. 1999 \pasj, 51, 321

\bibitem[Matsumoto et al.(2001)]{Matsumoto01}
Matsumoto, H., Tsuru, T.,  Koyama, K., Awaki, H., Canizares, C.R., Kawai, N., 
Matsushita, and S., Kawabe, R. 2001, \apj, 547, L25

\bibitem[McDowell(1985)]{McDowell}
McDowell, J., 1985, \mnras, 217, 77

\bibitem[Monet et al.(1998)]{Monet}
Monet, D., et al., 1998, The USNO-A2.0 Catalogue, (U.S. Naval Observatory, Washington D.C.)

\bibitem[Narayan \& Yi(1994)]{NarayanYi}
Narayan, R., \& Yi, I. 1994, \apj, 428, L13

\bibitem[Novikov \& Thorne(1973)]{Novikov}
Novikov, I.D. \& Thorne, K.S., 1973, in Black Holes, ed.\ C.\ DeWitt \&
B. DeWitt (New York: Gordon \& Breach)

\bibitem[Popov \& Prokhorov(1998)]{Popov}
Popov, S. B., \& Prokhorov, M. E., 1998, \aap, 331, 535

\bibitem[Ptak \& Griffith(1999)]{Ptak} 
Ptak, A. and Griffith, R. 1999 \apjl, 535, L85

\bibitem[Rees, Begelman \& Phinney(1982)]{Rees} 
Rees, M. J., Begelman, M. C., Blandford, R. D. \& Phinney, E. S. 1982,
\nat, 295, 17

\bibitem[Reimers et al.(1998)]{Reimers}
Reimers, D., Jordon, S., Beckmann, V., Christlieb, B. \& Wisotzki, L. 1998, \aap, 337, L13 

\bibitem[Revnivtsev \& Sunyaev(2001)]{Revnivtsev}
Revnivtsev, M. G., \& Sunyaev, R. A., 2001, astro-ph/0109036

\bibitem[Richards et al.(2001)]{Richards}
Richards, G. T., et al., 2001, \aj, 121, 2308

\bibitem[Schneider et al.(2002)]{Schneider}
Schneider, D. P., et al., 2002, \aj, 123, 567

\bibitem[Shakura \& Sunyaev(1973)]{Shakura}
Shakura, N. I., \& Sunyaev, R. A., 1973, \aap, 24, 337

\bibitem[Shapiro(1973)]{Shapiro}
Shapiro, S. L., 1973, \apj, 180, 531

\bibitem[Shapiro \& Teukolsky(1983)]{Shapiro2}
Shapiro, S. L., \& Teukolsky, S. A., 1983, Black Holes, White Dwarfs and Neutron Stars (New York: John Wiley \& Sons)

\bibitem[Shvartsman(1971)]{Shvartsman}
Shvartsman, V.F., 1971, Soviet Astron.---AJ, 15, 377


\bibitem[Stoughton et al.(2002)]{Stoughton}
Stoughton, C., et al., 2002, \aj, 123, 485

\bibitem[Taam \& Fryxell(1989)]{Taam}
Taam, R. E., \& Fryxell, B. A., 1989, \apj, 339, 297

\bibitem[Tanaka \& Lewin(1996)]{Tanaka}
Tanaka, Y., \& Lewin, W. H. G. 1996, in X-ray Binaries,
ed. W. H. G. Lewin, J. van Paradijs, \& E. P. J. van den Heuvel
(Cambridge: Cambridge University Press)

\bibitem[van Paradjis \& White(1995)]{VanParadjis}
van Paradjis, J. \& White, N. E., 1995, \apjl, 447, L33

\bibitem[Venkatesan, Olinto \& Truran(1999)]{Venkatesan}
Venkatesan, A., Olinto, A. V., \& Truran, J. W., 1999, \apj, 516, 863

\bibitem[V\'{e}ron-Cetty \& V\'eron(2001)]{Veron}
V\'{e}ron-Cetty, M.-P. \& V\'{e}ron, P. 2001, \aap, 374, 92

\bibitem[Voges et al.(1999)]{Voges}
Voges, W., et al. 1999, \aap, 349, 389

\bibitem[White \& van Paradjis(1996)]{White}
White, N. E. \& van Paradjis, J., 1996, \apjl, 473, L25

\bibitem[York et al.(2000)]{York}
York, D. G., et al. 2000, \aj, 120, 1579

\bibitem[Zeldovich \& Novikov(1971)]{Zeldovich}
Zeldovich, Ya.B., \& Novikov, I.D., 1971, Relativistic Astrophysics, Vol.\ 1
(Chicago: Univ.\ Chicago Press)

\end{thebibliography}
\end{document}